\documentclass[twocolumn,floatfix,showpacs,amsmath,amssymb,pra]{revtex4}
\usepackage{times}
\usepackage{graphicx}
\usepackage{dcolumn}
\usepackage{bm}

\begin{document}

\title{Photoassociation dynamics in a Bose-Einstein condensate}
\author{Thomas Gasenzer}
\thanks{email:T.Gasenzer@thphys.uni-heidelberg.de}
\affiliation{Institut f\"ur Theoretische Physik, Universit\"at Heidelberg, Philosophenweg 16, 69120 Heidelberg, Germany}
\date{\today}

\begin{abstract} 
\noindent
A dynamical many body theory of single color photoassociation in a Bose-Einstein condensate is presented.
The theory describes the time evolution of a condensed atomic ensemble under the influence of an arbitrarily varying near resonant laser pulse, which strongly modifies the binary scattering properties.
In particular, when considering situations with rapid variations and high light intensities the approach described in this article leads, in a consistent way, beyond standard mean field techniques.
This allows to address the question of limits to the photoassociation rate due to many body effects which has caused extensive discussions in the recent past.
Both, the possible loss rate of condensate atoms and the amount of stable ground state molecules achievable within a certain time are found to be stronger limited than according to mean field theory.
By systematically treating the dynamics of the connected Green's function for pair correlations the resonantly driven population of the excited molecular state as well as scattering into the continuum of non-condensed atomic states are taken into account.
A detailed analysis of the low energy stationary scattering properties of two atoms modified by the near resonant photoassociation laser, in particular of the dressed state spectrum of the relative motion prepares for the analysis of the many body dynamics.
The consequences of the finite lifetime of the resonantly coupled bound state are discussed in the two body as well as in the many body context.
Extending the two body description to scattering in a tight trap reveals the modifications to the near resonant adiabatic dressed levels caused by the decay of the excited molecular state.
\end{abstract}
\pacs{03.75.Nt, 03.75.Kk, 33.70.-w, 05.30.-d\hfill HD--THEP--04--25}
\maketitle
\section{Introduction}
\label{sec:intro}
Since the pioneering achievement of Bose-Einstein condensation in dilute alkali gases much effort has been made to extend this success to a wider class of atomic and molecular species.
In particular, the production of quantum degenerate ensembles of molecules \cite{GossLevi00Williams00} promises a wide spectrum of important applications ranging from ultra-precise molecular spectroscopy and cold collision studies \cite{Weiner99} to ``superchemical'' reactions \cite{Heinzen00} and the investigation of a possible crossover from a Bose-Einstein condensate (BEC) of molecules to Bardeen-Cooper-Schrieffer (BCS) correlated pairs in Fermi gases \cite{Randeria95}.
Laser cooling of molecules is essentially ruled out due to the complicated rovibrational molecular level structure \cite{GossLevi00Williams00}, such that alternative methods are in order when proceeding towards degenerate molecular gases.
The most promising procedure to date is photoassociation (PA) in ultracold, possibly degenerate atomic gases \cite{Julienne98,Weiner99}.
A photoassociative scattering resonance may occur when a conveniently tuned laser field induces a transition of a colliding atom pair to an electronically excited molecular, i.e.~an excimer bound state. 
PA in BECs has first been reported in Ref.~\cite{Wynar00,Gerton00} and recently been performed at considerably higher rates \cite{McKenzie02,Prodan03}.
Here, we present a dynamical many body theory which generalizes the approach in Refs.~\cite{KB02,KGB03,KGJB03} to photoassociation in a BEC and systematically takes into account effects of correlation functions beyond the condensate mean field.

On the broadway towards molecular BEC considerable success has been achieved recently in a number of exciting experiments employing Feshbach resonances.
Such resonances are formally analogous to PA resonances but they are not complicated by a short lifetime of the resonantly coupled molecular state.
Macroscopic amounts of ultracold diatomic molecules were assembled both from Bose-Einstein condensed atoms \cite{Donley02,Claussen03,Herbig03,Duerr03,Xu03} and in degenerate two-component gases of fermionic atoms \cite{Regal03,Strecker03,Cubizolles03,Jochim03a,Regal04a,Greiner03,Jochim03b,Zwierlein03,Bartenstein03,Regal04b,Zwierlein04,Bartenstein04}.

Despite these remarkable achievements PA in the degenerate regime remains at the top of the list of desirable aims. 
PA promises that one may use far developed laser techniques to form molecular quantum gases of arbitrarily modelled compounds.
Techniques available so far are limited due to insufficient overlap between molecular states in different scattering channels as well as by the rapid spontaneous decay of the intermediate excimer states.
This decay may be controlled by applying a second laser beam which drives the transition of the excited molecules to a stable bound state of atoms in the electronic ground state \cite{Julienne98,Heinzen00,Wynar00}, particularly in the form of Stimulated Raman Adiabatic Passage (STIRAP) \cite{STIRAP,STIRAPMBEC}.
Coherent control techniques which achieve optimized laser pulses \cite{CoherentControl} may further help to produce condensates of molecules in conveniently engineered states.
Theoretically, PA involving one or more laser modes has been studied in detail in the framework of two body scattering \cite{Fedichev96,Bohn96,Bohn97,Julienne98,Bohn99,Weiner99} as well as for quantum degenerate gases containing bosonic \cite{TheoryManybodyPAinBEC,Kostrun00,Goral01,Holland01,Javanainen02} and/or fermionic atoms \cite{TheoryManyBodyPAwithFermions}.
An important question which remains to be answered both experimentally and theoretically concerns limitations to achievable association rates due to many body effects \cite{Bohn99,Goral01,Holland01,Javanainen02,McKenzie02,Prodan03,Schloeder02}.

Early approaches to the many body dynamics of a BEC near a photoassociation resonance were based on mean field theory in the form of coupled Gross-Pitaevskii equations for atoms and molecules \cite{GPEsforMBEC}.
Beyond mean field level the effect of the pair correlation function has been taken into account \cite{Holland01,KGB03,Duine03,FBTheorybeyondMF}.
All of these approaches, including pure mean field models, are expected to satifactorily describe condensate loss and molecule formation \emph{in adiabatic crossings} of a Feshbach resonance, in which the magnetic field is \emph{varied linearly in time} \cite{KGG03,GKGJT04}.
It was shown, however, in Ref.~\cite{KGB03}, that a non-trivial variation of the magnetic field near such a resonance may lead to effects which are not accounted for by coupled Gross-Pitaevskii equations.

In view of these results a systematic treatment of the dynamics of higher order correlation functions seems is in order, which allows to describe strong coupling PA in a BEC and to identify phenomena which are distinctly different from those near a Feshbach resonance. 

The article is structured as follows:
Section \ref{sec:2bodytheory} is focused on the two body physics of low energetic atoms colliding in the presence of the near resonant laser mode.
The dynamical evolution of the many body system starting as a trapped Bose-Einstein condensate is studied in Section \ref{sec:manybodytheory}.  
Section \ref{sec:concl} contains a summary of the results and draws our conclusions.
An appendix contains explicit descriptions of the techniques used to describe the two body collisions and the many body dynamics as relevant for this work.

In Section \ref{sec:2bodytheory} we define the two-channel picture of resonant single color photoassociative scattering of a pair of $^{23}$Na atoms as in the recent experiment of McKenzie {\it et al.} \cite{McKenzie02}. 
The scattering length approach to the laser assisted atomic collisions \cite{Fedichev96,Bohn96} is introduced, and the relevant parameters characterizing the resonance are identified.
These parameters define a minimal model Hamiltonian which suffices to describe all the relevant low energy scattering properties. 
Using this Hamiltonian we focus on the set of atomic pair states dressed by the laser with a contribution from the excimer bound state.
To illustrate the effects of the finite lifetime of this bound state the dressed state spectrum is studied for pairs confined in a tight harmonic trap for which
a convenient modification of the minimal model Hamiltonian is developed.
In this way we could study the existence of quasi-bound states above threshold which are a consequence of the excimer state decay.
We then consider the components of the dressed state closest to threshold corresponding to the background channel and the admixed excimer state.
We show that for high light intensities the near resonant dressed state resembles a bound state in the background channel which, as discussed in Section \ref{sec:manybodytheory}, has a decisive impact on the corresponding many body dynamics and on possible single color PA rates.

In Section \ref{sec:manybodytheory} we extend the many body dynamical theory developed in Refs.~\cite{KB02,KGB03,KGJB03,KGG03,GKGJT04} to single color PA in a BEC.
We discuss in detail the incorporation of the resonant two body physics into the many body formalism and the r\^{o}le of effects beyond the Gross-Pitaevskii mean field level.
Since the theory exactly respects atom number conservation at all times it also allows us to study the overall loss of atoms due to the spontaneous decay of the excited molecular state.
As a result we obtain limits to both, the achievable condensate loss rate and the maximum rate at which ground state molecules may be formed by spontaneous electronic deexcitation.
The different origins of these limits will be discussed.
Our results show that the rate limit for condensate loss predicted in Ref.~\cite{Javanainen02} is exceeded at high light intensities.
Nevertheless, we find that the maximum amount of atoms lost within a certain time interval is reduced compared to the value derived from a naive Gross-Pitaevskii approach.
This limit, however has not yet been reached at the intensities in the experiment in Ref.~\cite{McKenzie02}. 
We finally present examples for the time evolution of the density profile of a trapped ensemble under high light intensity PA which exhibit strong collective excitations of the atomic cloud.
\begin{figure}[tb]
\begin{center}
\includegraphics[width=0.45\textwidth]{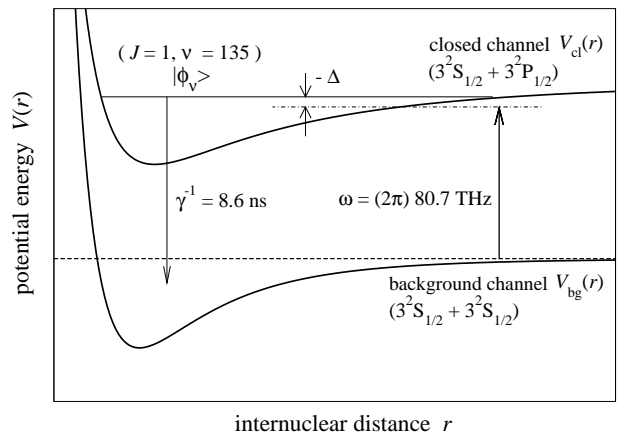}
\end{center}
\vspace*{-3ex}
\caption{
Schematic interaction potentials of $^{23}$Na pairs coupled by a single laser mode.
The condensate atoms collide with a relative energy close to the background channel threshold indicated by a dashed line.
They are excited by a laser with frequency $\omega=(2\pi)\,80.7\,$THz to the $(J=1,\nu=135)$ rovibrational bound state in the closed channel and spontaneously decay by photon emission to ground state molecules as well as free atom pairs, with a time constant of $\gamma^{-1}=8.6\,$ns.
$\Delta=\omega-\mathrm{Re}E_\nu/\hbar$ denotes the detuning of the laser. 
}
\label{fig:multichannelsketch}
\end{figure}

\section{Laser assisted resonant atomic collisions}
\label{sec:2bodytheory}
Consider two atoms which are colliding in the presence of a light field from a laser source.
PA of the pair is generally understood as its light induced binding to a diatomic molecule.
We would like to choose a somewhat more general definition which includes any laser controlled modification of the collisional interaction properties of the atoms.
A key finding then is that PA allows to modify the low energy scattering properties of atoms in an ultracold vapour in a way very similar to what is known as a Feshbach resonance \cite{Fedichev96,Bohn96}.
Although very different physically much of the theoretical formalism used for one kind of scattering resonance, photoassociative or Feshbach resonant, may also be applied to the other.

The main aim of the work presented in this article is to adapt the microscopic dynamical many body theory of Refs.~\cite{KB02,KGB03} to photoassociation of Bose-Einstein condensed atoms.
In this theory the binary interactions are incorporated in a form in which they are treated in quantum mechanical scattering theory.
We shall therefore introduce, in this section, the quantum mechanical description of the low energy binary collisions in the presence of the laser field.
The photoassociative collisions are modelled in a two channel picture which accounts for asyptotically free atom pairs with a certain relative energy coupled by the laser to a closed channel molecular bound state, cf.~Fig.~\ref{fig:multichannelsketch}.
We follow the approach of Refs.~\cite{Fedichev96,Bohn96} in characterizing the low energy scattering in terms of a laser modified $s$-wave scattering length.
Due to its relatively strong dipole moment the closed channel state is short lived such that the physical situation is substantially modified as compared to the Feshbach case.
We shall study in detail the consequences of this decay for the spectrum of dressed states of an atom pair in the vicinity of the resonance.

\subsection{Photoassociation of $^{23}$Na atoms}
\label{sec:sodium}
In this article we focus on single color photoassociation in a trapped condensate of $^{23}$Na atoms.
Related experimental results have been presented by McKenzie {\it et al.} in Ref.~\cite{McKenzie02}.
In this experiment the condensed atoms are initially in the $3^2$S$_{1/2}$, $|F=1,m_F=-1\rangle$ hyperfine state, and their binary collisions in the absence of laser light are described by a mixture of the X$^1\Sigma_g^+$ singlet and the a$^3\Sigma^+_u$ triplet potentials \cite{Julienne98}.
Atom pairs in the condensate which initially are moving freely on the nearly degenerate asymptote of these ground state potentials are coupled by a laser with frequency $\omega=(2\pi)\,80.7\,$THz to the ($J=1$, $\nu=135$) rovibrational level of the A$^1\Sigma_u^+$ Na$_2$ molecular potential (Fig.~\ref{fig:multichannelsketch}).
This excited molecular state $|\phi_\nu\rangle$ has a lifetime of about $8.6\,$ns which corresponds to a linewidth of $\gamma=(2\pi)\,18.5\,$MHz being nearly twice the atomic linewidth.
The spontaneous decay of $|\phi_\nu\rangle$ to hot atom pairs or ground state molecules is taken into account in the imaginary part of the energy $E_\nu=\hbar(\omega_\nu-i\gamma/2)$ of $|\phi_\nu\rangle$.
\begin{figure}[tb]
\begin{center}
\includegraphics[width=0.45\textwidth]{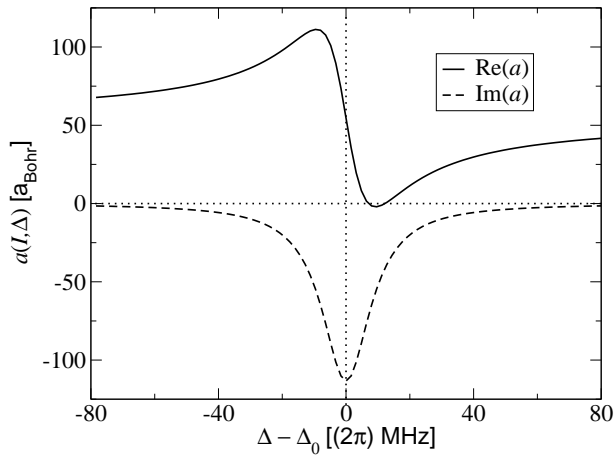}
\end{center}
\vspace*{-3ex}
\caption{
Complex $s$-wave scattering length $a(I,\Delta)$ describing the low energy scattering of $^{23}$Na atoms, in the vicinity of a scattering resonance induced by single color photoassociation,  for $I=1\,$kW/cm$^2$, as a function of the detuning $\Delta$ (see Fig.~\ref{fig:multichannelsketch}).
}
\label{fig:aofIandomega}
\end{figure}
If the intensity of the laser light is kept fixed, the tuning of the frequency through the transition leads to a resonant variation of the $s$-wave scattering length \cite{Fedichev96,Bohn96,Bohn99}
\begin{align}
  \label{eq:aofIandomega}
  a(I,\Delta) 
  = a_\mathrm{bg}\left(1-\frac{\delta(I)}{\Delta_0(I)-\Delta-i\gamma/2}
                 \right),
\end{align}
as shown in Fig.~\ref{fig:aofIandomega}.
The description of low energy PA scattering in terms of a scattering length indicates the formal similarity to a Feshbach resonance. 
However, due to the finite lifetime of $|\phi_\nu\rangle$ the PA scattering length is complex and its modulus does not diverge on resonance.
$\Delta=\omega-\omega_\nu$ is the detuning of the laser from the frequency difference between the molecular level $\nu$ and the ground state asymptote.
The resonance is shifted by $\Delta_0(I)=\omega_0(I)-\omega_\nu$.
On resonance, $\omega=\omega_0(I)$, the real part of the scattering length is equal to the background scattering length $a_\mathrm{bg}$ which characterizes the interactions in absence of the laser.
The width of the resonance is parametrized by $\delta(I)$.
Both, $\delta(I)$ and the shift $\Delta_0(I)$ depend on the light intensity $I$ and, for low intensities, were found to vary linearly with $I$ \cite{McKenzie02} as predicted by perturbation theory \cite{Bohn99}.
The scattering length shown in Fig.~\ref{fig:aofIandomega} was determined with Eq.~(\ref{eq:aofIandomega}) using the values $\delta(I)=(2\pi)\,38.4\,$MHz, and $\Delta_0(I)=(2\pi)\,164\,$MHz, for $I=1.0\,$kW/cm$^2$ \cite{McKenzie02}.

\subsection{Two channel model of resonant two body scattering}
\label{sec:multichannel_2b}
Figure \ref{fig:multichannelsketch} shows the general process we aim to describe.
Two ultracold condensate atoms with a relative kinetic energy close to the free-bound threshold of the potential $V_\mathrm{bg}$ interact with each other.
In the absence of the laser light their collisional interaction is described by  $V_\mathrm{bg}$.
Since the total energy of the atom pair lies above threshold this background scattering channel is termed an open channel. 
For simplicity we consider only a single open channel potential, e.g.~, in the case of sodium atoms colliding in the $|F=1,m_F=-1\rangle$ state, a potential which reproduces the low energy scattering properties of condensed sodium atoms as measured in experiment or calculated theoretically using more elaborate multichannel techniques \cite{Julienne98}.
We neglect the explicit notation of the internal state, of fine and hyperfine quantum numbers, in order to focus on the structure of the light induced coupling of the states.

The potential which describes the interactions of atoms in the S$_{1/2}$ ground state with atoms in the excited P$_{1/2}$ state is denoted as $V_\mathrm{cl}$.
The difference of the threshold energies of $V_\mathrm{cl}$ and $V_\mathrm{bg}$ equals the atomic line frequency and, since this is in practice larger than $\omega$, $V_\mathrm{cl}$ is termed a closed channel potential for the near zero energy condensate atoms.

We assume the laser to be described as a classical plane wave with wave vector $\mathbf{k}$ and frequency $\omega=|\mathbf{k}|c$.
The light gives rise to a radiative coupling $W(\mathbf{x},t)=\hbar\Omega(\mathbf{x},t)$ expressed in terms of a time dependent Rabi frequency $\Omega(\mathbf{x},t)=\Omega_0(t)\cos[\mathbf{k}\mathbf{x}-\omega t]$.
Here, we have included the temporal variation of the envelope of the light field, i.e. the time dependence of its intensity $I(t)$ in the Rabi frequency $\Omega_0(t)$.

The Hamiltonian operator describing the binary interactions in the two channel picture may now be written as $H_\mathrm{2B}=H_0+V_\mathrm{2B}$, with
\begin{align}
 \label{eq:H2B0}
  H_0
  &=\ T_\mathrm{kin}\big[
   |\mathrm{bg}\rangle\langle\mathrm{bg}|
  +|\mathrm{cl}\rangle\langle\mathrm{cl}|\Big]
  \\
 \label{eq:V2B}
  V_\mathrm{2B}
  &=\ (|\mathrm{bg}\rangle,|\mathrm{cl}\rangle)
      \left(\begin{array}{cc}
      V_\mathrm{bg} & 
      W \\
      W  & 
      V_\mathrm{cl}
      \end{array}\right)
      \left(\!\!\begin{array}{c}
      \langle\mathrm{bg}|\\ 
      \langle\mathrm{cl}|
      \end{array}\!\!\right).
\end{align}
Here, $T_\mathrm{kin}=-\hbar^2\nabla^2/m$ is the relative kinetic energy. $|\mathrm{bg}\rangle$ and $|\mathrm{cl}\rangle$ are the superposition of products of atomic states associated with the open and closed channels, respectively.
In Eq.~(\ref{eq:H2B0}) the trapping potential may be neglected since it does not vary on the scale of the binary interactions.

The range of relative kinetic energies in ultracold collisions is small compared to the typical spacing between vibrational states.
Throughout this work we will therefore consider only situations where the laser couples to a single well defined closed channel bound state $|\phi_\nu\rangle$. 
The closed channel Hamiltonian $H_{\rm cl}=T_\mathrm{kin}+V_{\rm cl}$ can then be replaced by the operator $E_\nu P_\nu$, with the quasi projector $P_\nu=|\phi_\nu\rangle\langle\tilde\phi_\nu|$ onto the excimer state, $H_{\rm cl}|\phi_\nu\rangle=E_\nu|\phi_\nu\rangle$.
Because of the spontaneous decay of the $3^2$P$_{1/2}$ excited atoms the $\langle\tilde\phi_\nu|$ are {\it left} eigenstates of the non-Hermitian operator $H_{\rm cl}$ to the {\it same} complex eigenvalue $E_\nu$, $\langle\tilde\phi_\nu|H_{\rm cl}=\langle\tilde\phi_\nu|E_\nu$. 
Cf.~Appendix \ref{app:EigenstatesNHOP} for details.

Within the subspace spanned by the open channel states and the single closed channel state $|\phi_\nu\rangle$ the Hamiltonian reads:
\begin{align}
 \label{eq:H2B0nu}
  H_0
  &=\ T_\mathrm{kin}
   |\mathrm{bg}\rangle\langle\mathrm{bg}|
  +|\phi_\nu,\mathrm{cl}\rangle E_\nu \langle\tilde\phi_\nu,\mathrm{cl}|
  \\
 \label{eq:V2Bnu}
  V_\mathrm{2B}
  &=\ (|\mathrm{bg}\rangle,|\mathrm{cl}\rangle)
      \left(\begin{array}{cc}
      V_\mathrm{bg} & 
      WP_\nu \\
      P_\nu W  & 
      0
      \end{array}\right)
      \left(\!\!\begin{array}{c}
      \langle\mathrm{bg}|\\ 
      \langle\mathrm{cl}|
      \end{array}\!\!\right).
\end{align}
For the following discussion of the dressed state spectrum of the Hamiltonian it will be convenient to go to an interaction picture, where, in the rotating wave approximation, the phase $\exp[-i\omega t]$ of $W(\mathbf{x},t)$ is removed and the closed channel frequency $\omega_\nu=\mathrm{Re}E_\nu/\hbar$ in Eq.~(\ref{eq:H2B0nu}) is replaced by $-\Delta=\omega_\nu-\omega$.

\subsection{Two channel dressed states}
\label{sec:2chdressed}
It is well known that the energy of the uppermost vibrational bound state of two interacting particles determines the singularities of the $s$-wave scattering length.

A Feshbach resonance may be induced through Zeeman shifting atomic hyperfine levels and thus different scattering channels relative to each other by applying an external magnetic field.
The channel states are coupled to each other through internal atomic forces.
This leads to a modification of the bound state spectrum of the atom pair.
In contrast to this, the external electromagnetic field of a PA laser transfers energy to the atom pair.
As a result, the width and the shift of the scattering resonance depend on the intensity of the light.

The two body state resulting from the modification by the laser field is called a dressed state.
Moreover, 
since the lifetime of the closed channel state is very short
the population of the dressed states may not be measured in the same way as that of the bound states in recent Feshbach crossing experiments \cite{Regal03,Herbig03,Duerr03,Strecker03,Cubizolles03,Jochim03a,Zwierlein03}, viz.~by reversing the association process and imaging the atoms.
Stable molecules are rather obtained as bound states of the ground state potential $V_\mathrm{bg}$ when the closed channel molecule spontaneously emits a photon or when the transition is induced, e.g., by illuminating the sample with one or more additional lasers.
Such processes of Raman PA \cite{Julienne98}, possibly in the form of STIRAP \cite{STIRAPMBEC} or with optimized laser pulses (Coherent Control) \cite{CoherentControl} are beyond the scope of this article, but may be described in a straightforward extension of the formalism presented here.

Nevertheless, the dressed state picture is formally equivalent to the description of the modified bound state spectrum near a Feshbach resonance.
In the interaction picture the `energy' of the closed channel state compared to the open channel threshold is given by the laser detuning which one may, e.g., ramp through the resonance in much the same way as it is done in the case of adiabatic molecule association at a Feshbach resonance.

For these reasons we shall in this subsection discuss in detail the spectrum of dressed states of interacting atoms pairs in the framework of the two-channel picture introduced in Sect.~\ref{sec:multichannel_2b}.
After a discussion of the low energy scattering in the background channel we will derive the continuum of scattering states as well as the discrete part of the dressed state spectrum close to the PA resonance.
In Section \ref{sec:resboundstates} we will describe the particular consequences of the spontaneous decay of the closed channel state for the structure of the dressed state spectrum.

\subsubsection{Background scattering}
\label{sec:bgscatt}
As was discussed in detail in Refs.~\cite{KGB03,GKGJT04} the binding energy of the highest vibrational state of $V_\mathrm{bg}$ is determined, to an excellent approximation, by $a_\mathrm{bg}$ and the long range $r$-dependence of $V_\mathrm{bg}(r)$.
For the case of ground state alkali atoms, this is dominated by the van der Waals interaction, $V_\mathrm{bg}(r)\sim-C_6/r^6$, setting a length scale,
$l_\mathrm{vdW}=(mC_6/8\hbar^2)^{1/4}$ which is much smaller than the thermal de Broglie wave length in typical ultracold collisions.
Using semiclassical arguments it was shown in Ref.~\cite{Gribakin93} that the first correction to the energy of the highest vibrational state is given through
\begin{align}
  \label{eq:Ebgbofabar}
  E_{-1}
  &=\ -\frac{\hbar^2}{m(a_\mathrm{bg}-\bar{a})^2},
\end{align}
i.e.~by the usual formula for the energy in terms of the scattering length which, however, is shifted due to the van der Waals tail by the mean scattering length  $\bar{a}=l_\mathrm{vdW}\Gamma(3/4)/(\sqrt{2}\Gamma(5/4))\simeq0.956\,l_\mathrm{vdW}$. 
For the case of sodium considered here one has $C_6=1561\,$a.u.~\cite{Kharchenko97} ($1\,$a.u.~$=0.095734\cdot10^{-24}\,$J nm$^6$) and thus $l_\mathrm{vdW}=45.0\,a_\mathrm{Bohr}$ ($a_\mathrm{Bohr}=0.052918$ nm), and, with $a_\mathrm{bg}=54.6\,a_{\rm Bohr}$ \cite{Mies00}, a binding energy of $E_{-1}=(2\pi)1164\,$MHz$\,\hbar$.

The scattering of sodium atoms in the $|F=1,m_F=-1\rangle$ state is described by a mixture of the X$^1\Sigma_g^+$ singlet and a$^3\Sigma^+_u$ triplet potentials with scattering lengths for which Ref.~\cite{Mies00} gives values of $a_{S=0}=20.3\,a_\mathrm{Bohr}$ and $a_{S=1}=63.9\,a_\mathrm{Bohr}$, respectively.
The energies of the last bound states of the singlet and triplet potentials have been experimentally determined as $E_{-1}(S=0)=-318(21)\,$MHz$\,h$ and $E_{-1}(S=1)=-270(36)\,$MHz$\,h$, respectively \cite{Elbs99,Samuelis00}.
Hence the formula (\ref{eq:Ebgbofabar}) is not applicable for the sodium potentials.
This is due to the fact that the binding energies are too large such that the mean scattering length $\bar{a}=43\,a_\mathrm{Bohr}$ is not a small correction to the background scattering length $a_\mathrm{bg}=54.6\,a_{\rm Bohr}$ \cite{Gribakin93,fn:Rb85case}.

However, it was shown in Ref.~\cite{Gao98} that the energies of the uppermost bound state of the singlet and triplet potentials are excellently determined by the scattering length and the van der Waals coefficient $C_6$ alone.
At zero collision energy the scattering length $a_\mathrm{bg}$ summarizes all the unresolved details of $V_\mathrm{bg}(r)$ into a single length scale.
Atoms interacting at finite collision energies resolve details of the potential, and the long range behaviour quantified by the van der Waals length determines the leading contribution.
Any model of the background scattering potential recovers the binding energy of the highest vibrational state and the low energy scattering properties if it properly accounts for the parameters $a_\mathrm{bg}$ and $C_6$.

In this work we shall use the experimentally determined value of $E_{-1}=-318\,$MHz$\,h$ to model the long range behaviour of the background potential.
We introduce a minimal background potential in Appendix \ref{app:seppot} which is related to a separable approximation of the full interaction potential.

\subsubsection{Dressed state spectrum: Continuum part}
\label{sec:contstates}
Assuming that the spectrum and states associated with the background potential are known, we shall, in this section, derive the continuum of scattering states of the full two channel Hamiltonian, Eqs.~(\ref{eq:H2B0nu}), (\ref{eq:V2Bnu}).
Moreover we will deduce the dependence of the scattering length $a(I,\omega)$ on the relevant quantities contained in the Hamiltonian.
This will enable us to describe the full resonant two channel scattering properties at low collision energies by a minimal set of five parameters.

The two channel eigenstates of the Hamiltonian, Eqs.~(\ref{eq:H2B0nu}), (\ref{eq:V2Bnu}), may be written in terms of their open and closed channel components as $|\mathrm{bg}\rangle|\phi^\mathrm{bg}\rangle+|\mathrm{cl}\rangle|\phi^\mathrm{cl}\rangle$.
The two components are solutions of the stationary two channel Schr\"odinger equation, which, in the interaction picture introduced in Sect.~\ref{sec:multichannel_2b}, reads
\begin{align}
  \label{eq:TwoChSSE}
  \left(\begin{array}{cc}
   H_\mathrm{bg}
  &WP_\nu \\
   P_\nu W
  &H_\mathrm{cl}
  \end{array}\right)
  \left(\begin{array}{c}
   |\phi^\mathrm{bg}\rangle \\
   |\phi^\mathrm{cl}\rangle
  \end{array}\right)
  &= E
  \left(\begin{array}{c}
   |\phi^\mathrm{bg}\rangle \\
   |\phi^\mathrm{cl}\rangle
  \end{array}\right),
\end{align}
where $H_\mathrm{bg}=T_\mathrm{kin}+V_\mathrm{bg}$ and $H_\mathrm{cl}=(E_\nu-\hbar\omega) P_\nu=-\hbar(\Delta+i\gamma/2)P_\nu$, with $P_\nu=|\phi_\nu\rangle\langle\tilde\phi_\nu|$.

As discussed in Sect.~\ref{sec:multichannel_2b} the laser is assumed to couple the ground state atom pair to a single unstable bound state in the closed channel potential.
As a consequence, the closed channel component of the scattering state vanishes at large interatomic distances,
\begin{align}
  \label{eq:asymptotclScattWF}
  \phi_\mathbf{p}^\mathrm{cl}(\mathbf{r})
  &\underset{r\rightarrow\infty}{\sim}
  0.
\end{align}
Assuming that there is no other decay channel, the continuum of scattering states then consists of all solutions of Eq.~(\ref{eq:TwoChSSE}) with \emph{real} positive energy $E=p^2/m$, corresponding to a relative momentum $p=|\mathbf{p}|$ of the atoms colliding in the open channel.
An imaginary part of $E$ is excluded since the atoms are in the ground state and thus stable at large distances $r$ where the coupling to the closed channel resonance state is negligible.
The background scattering states at large $r$ are chosen to obey the well known expression in terms of an incoming plane wave, an outgoing spherical wave, and a scattering amplitude $f(\vartheta,p)$ which depends on $p$ and the angle $\vartheta$ between $\mathbf{p}$ and the relative position $\mathbf{r}$ of the outgoing atoms,
\begin{align}
  \label{eq:asymptotbgScattWF}
  \phi_\mathbf{p}^\mathrm{bg}(\mathbf{r})
  &\underset{r\rightarrow\infty}{\sim}
  (2\pi\hbar)^{-3/2}\left[e^{i\mathbf{p}\mathbf{r}/\hbar}
  +f(\vartheta,p)\frac{e^{ipr}}{r}\right].
\end{align}

The Schr\"odinger equation (\ref{eq:TwoChSSE}) may then be expressed in terms of the Green's functions
\begin{align}
  \label{eq:Gbgofz}
  G_\mathrm{bg}(z)
  &= \left(z-H_\mathrm{bg}\right)^{-1},
  \\
  \label{eq:Gclofz}
  G_\mathrm{cl}(z,\Delta)
  &= \frac{P_\nu}{z+\hbar(\Delta+i\gamma/2)},
\end{align}
which depend on a complex energy parameter $z$ \cite{fn:poleapprox}.
For the scattering states the equation reads:
\begin{align}
  \label{eq:IntTwoChSSE}
  \left(\begin{array}{c}
   |\phi^\mathrm{bg}_\mathbf{p}\rangle \\
   |\phi^\mathrm{cl}_\mathbf{p}\rangle
  \end{array}\right)
  &=
  \left(\begin{array}{c}
   |\phi^{(+)}_\mathbf{p}\rangle \\
   0
  \end{array}\right)
  +
  \left(\begin{array}{c}
   G_\mathrm{bg}(E+i0)WP_\nu|\phi^\mathrm{cl}_\mathbf{p}\rangle \\
   G_\mathrm{cl}(E,\Delta)P_\nu W|\phi^\mathrm{bg}_\mathbf{p}\rangle
  \end{array}\right).
\end{align}
The imaginary part `$i0$' of the argument of $G_\mathrm{bg}$ indicates that the Green's function is evaluated in the limit of approaching the energy $E$ from the upper half of the complex plane, ensuring that the wave function $\phi^\mathrm{bg}_\mathbf{p}(\mathbf{r})$ has the asymptotic behavior given in Eq.~(\ref{eq:asymptotbgScattWF}).

In Eq.~(\ref{eq:IntTwoChSSE}), $|\phi_\mathbf{p}^{(+)}\rangle$ denotes the positive energy solution of the homogeneous Schr\"odinger equation, i.e.~Eq.~(\ref{eq:TwoChSSE}) with $W$ set to zero, behaving at large interatomic distances as
\begin{align}
  \label{eq:homasymptotbgScattWF}
  \phi_\mathbf{p}^\mathrm{(+)}(\mathbf{r})
  &\underset{r\rightarrow\infty}{\sim}
  (2\pi\hbar)^{-3/2}\left[e^{i\mathbf{p}\mathbf{r}/\hbar}
  +f_\mathrm{bg}(\vartheta,p)\frac{e^{ipr}}{r}\right].
\end{align}

Given the pole approximated closed channel Green's function $G_\mathrm{cl}(E,\Delta)$, Eq.~(\ref{eq:Gclofz}), the closed channel component of the scattering state, Eq.~(\ref{eq:IntTwoChSSE}), is proportional to the resonant state $|\phi_\nu\rangle$.
Thus the scattering state may be written as
\begin{align}
  \label{eq:ScattState}
  \left(\!\begin{array}{c}
   |\phi^\mathrm{bg}_\mathbf{p}\rangle \\
   |\phi^\mathrm{cl}_\mathbf{p}\rangle
  \end{array}\!\right)
  &=
  \left(\!\begin{array}{c}
   |\phi^{(+)}_\mathbf{p}\rangle \\
   0
  \end{array}\!\right)
  +
  \left(\!\begin{array}{c}
   G_\mathrm{bg}(E+i0)W|\phi_\nu\rangle \\
   |\phi_\nu\rangle
  \end{array}\!\right) A(E,\Delta),
\end{align}
with the amplitude
\begin{align}
  \label{eq:Aitophipbg}
  A(E,\Delta)
  &= \frac{\langle\tilde\phi_\nu|W|\phi^\mathrm{bg}_\mathbf{p}\rangle}
          {E+\hbar(\Delta+i\gamma/2)}
\end{align}
which from the background channel component of Eq.~(\ref{eq:ScattState}), is determined as
\begin{align}
  \label{eq:Aitophipplus}
  A(E,\Delta)
  &= \frac{\langle\tilde\phi_\nu|W|\phi^{(+)}_\mathbf{p}\rangle}
          {E+\hbar(\Delta+i\gamma/2)
	   -\langle\tilde\phi_\nu|WG_\mathrm{bg}(E+i0)W|\phi_\nu\rangle}.  
\end{align}
In summary: Given the full positive energy spectrum $|\phi^{(+)}_\mathbf{p}\rangle$ of eigenstates of $H_\mathrm{bg}$ as well as the resonantly coupled closed channel state $|\phi_\nu\rangle$, Eqs.~(\ref{eq:ScattState}) and (\ref{eq:Aitophipplus}) represent the exact solution of the two channel Schr\"odinger equation (\ref{eq:TwoChSSE}).
Under the assumption that the excitation of the colliding atom pair is restricted to the state $|\phi_\nu\rangle$ this pole approximated solution of the full two channel Schr\"odinger equation becomes exact.

We shall now derive formula (\ref{eq:aofIandomega}) for the scattering length near resonance.
The $s$-wave scattering length in the open scattering channel is defined as the negative of the zero momentum limit of the scattering amplitude $f(\vartheta,p)$ which is equivalent to the expression
\begin{align}
  \label{eq:aitoT}
  a
  &= \lim_{p\to0}(2\pi\hbar)^3\frac{m}{4\pi\hbar^2} 
     \langle\mathbf{p},\mathrm{bg}|T_\mathrm{2B}(p^2/m)
     |\mathbf{p},\mathrm{bg}\rangle
\end{align}
in terms of the on shell $T$-matrix element between plane wave states with momentum $\mathbf{p}$ in the background channel.
Using Eqs.~(\ref{eq:V2Bnu}), (\ref{eq:ScattState}), (\ref{eq:Aitophipplus}), and the definition $\langle\mathbf{p}|T_\mathrm{2B}(p^2/m)|\mathbf{p}\rangle=\langle\mathbf{p}|V_\mathrm{2B}|\phi^\mathrm{2ch}_\mathbf{p}\rangle$ of the diagonal plane wave $T$-matrix elements in terms of the potential $V_\mathrm{2B}$, Eq.~(\ref{eq:V2Bnu}), and a two channel scattering state $|\phi^\mathrm{2ch}_\mathbf{p}\rangle$ as given in Eq.~(\ref{eq:ScattState}), we find:
\begin{align}
  \label{eq:aitoMatrElts}
  a
  &= (2\pi\hbar)^3\frac{m}{4\pi\hbar^2} 
     \langle0|V_\mathrm{bg}|\phi^\mathrm{bg}_0\rangle
    +\langle0|WP_\nu|\phi^\mathrm{cl}_0\rangle
  \nonumber\\
  &= a_\mathrm{bg}
    -\frac{\frac{m}{4\pi\hbar^2}(2\pi\hbar)^3
           \langle\phi^{(+)}_0|WP_\nu W|\phi^{(+)}_0\rangle}
	  {\langle\tilde\phi_\nu|WG_\mathrm{bg}(0)W|\phi_\nu\rangle
	   -\hbar(\Delta+i\gamma/2)}.       
\end{align}
This reproduces the resonance formula for the scattering length given in Eq.~(\ref{eq:aofIandomega}) if one identifies
\begin{align}
  \label{eq:deltaitoME}
  \delta(I)
  &= \frac{m\hbar^{-1}}{4\pi\hbar^2a_\mathrm{bg}}(2\pi\hbar)^3
     \langle\phi^{(+)}_0|W(I)P_\nu W(I)|\phi^{(+)}_0\rangle,
  \\
  \label{eq:shiftitoME}
  \Delta_0(I)
  &= \hbar^{-1}
     \langle\tilde\phi_\nu|W(I)G_\mathrm{bg}(0)W(I)|\phi_\nu\rangle,
\end{align}
where we explicitly denote the dependence on the light intensity $I$.

Knowledge of the dipole matrix elements and the Franck-Condon factors would allow, on the basis of the known background scattering properties, to calculate the shift $\Delta_0$ and widths of the photoassociation resonance from the matrix elements in Eqs.~(\ref{eq:deltaitoME}), (\ref{eq:shiftitoME}) (cf., e.g.~Ref.~\cite{Simoni02}).
In this work we will however assume that the shift and width are quantities known from experiment over a certain range of light intensities $I$, and that they depend, in this range, linearly on $I$.
The experiment in Ref.~\cite{McKenzie02} confirms this linear dependence for the range of applicable intensities $I$ applied, and we will assume that the linear behaviour also extends to somewhat larger intensities.

In analogy to the minimal model potential describing the low energy scattering properties of the background channel interactions we shall introduce a characterization of the coupling to the closed channel state in terms of two parameters.
These two parameters measure the coupling strength and the range of interatomic distances where the coupling occurs and are fixed through Eqs.~(\ref{eq:deltaitoME}) and (\ref{eq:shiftitoME}) in terms of the measureable resonance width and shift.
Cf.~Appendix \ref{app:seppot} for details of how to choose an ansatz convenient for practical computations.

In summary, together with the two background parameters of the minimal model, and the decay width $\gamma$ of $|\phi_\nu\rangle$, they form a minimal set of five parameters which fully describe the low energy two body scattering of atoms under the influence of the near resonant photoassociation laser.
In giving explicit expressions for matrix elements of the Hamiltonian operator this minimal parametrization represents a powerful tool when solving the many body dynamical equations which will be presented in Section \ref{sec:manybodytheory}.

\subsubsection{Dressed state spectrum: Discrete part}
\label{sec:boundstates}
The discrete part of the spectrum of the two body Hamiltonian consists of dressed states with in general complex eigenenergies since the corresponding spatial wave functions vanish at large interatomic distances.
Similarly to Eq.~(\ref{eq:IntTwoChSSE}) the Schr\"odinger equation may be written in integral form,
\begin{align}
  \label{eq:IntTwoChSSEdrSt}
  \left(\begin{array}{c}
   |\phi^\mathrm{bg}_\mathrm{d}\rangle \\
   |\phi^\mathrm{cl}_\mathrm{d}\rangle
  \end{array}\right)
  &=
  \left(\begin{array}{c}
   G_\mathrm{bg}(E_d)WP_\nu|\phi^\mathrm{cl}_\mathrm{d}\rangle \\
   G_\mathrm{cl}(E_d,\Delta)P_\nu W|\phi^\mathrm{bg}_\mathrm{d}\rangle
  \end{array}\right),
\end{align}
and has, taking into account Eq.~(\ref{eq:Gclofz}), the normalized solution:
\begin{align}
  \label{eq:DressedState}
  \left(\begin{array}{c}
   |\phi^\mathrm{bg}_\mathrm{d}\rangle \\
   |\phi^\mathrm{cl}_\mathrm{d}\rangle
  \end{array}\right)
  &=
  \frac{1}{{\cal N}_d}
  \left(\begin{array}{c}
   G_\mathrm{bg}(E_d)W|\phi_\nu\rangle \\
   |\phi_\nu\rangle
  \end{array}\right).
\end{align}
$|\phi_\nu\rangle$ has by assumption unit norm such that the overall normalization constant is
\begin{align}
  \label{eq:NormDressed}
  {\cal N}_d
  &= \left[1+
     \langle\phi_\nu|W^\dagger G_\mathrm{bg}(E_\mathrm{d})^\dagger 
     G_\mathrm{bg}(E_\mathrm{d})W|\phi_\nu\rangle\right]^{1/2},
\end{align}
and the complex dressed state energy is determined by the condition
\begin{align}
  \label{eq:EnergyDressed}
  E_\mathrm{d}
  &= \langle\tilde\phi_\nu|WG_\mathrm{bg}(E_\mathrm{d})W|\phi_\nu\rangle
	   -\hbar(\Delta+i\gamma/2).
\end{align}

The states given through Eqs.~(\ref{eq:DressedState}) and (\ref{eq:EnergyDressed}) are \emph{right} eigenstates of the non-Hermitian Hamiltonian $H_\mathrm{2B}$, which are, in general, \emph{not} pairwise orthogonal to each other and to the scattering states given in Eq.~(\ref{eq:ScattState}) (Cf.~Appendix \ref{app:EigenstatesNHOP}).

\subsection{Properties of the dressed states near resonance}
\label{sec:resboundstates}
\subsubsection{Adiabatic energy levels: homogeneous case}
\label{sec:adiablevelsHom}
\begin{figure}[tb]
\begin{center}
\includegraphics[width=0.45\textwidth]{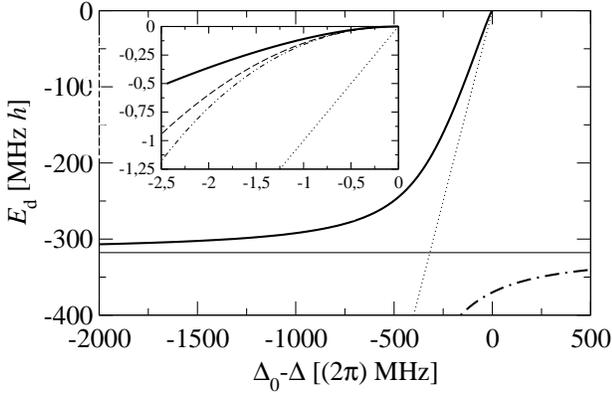}
\end{center}
\vspace*{-3ex}
\caption{
Energy $E_\mathrm{d}$ of the uppermost bound state of two sodium atoms for the hypothetical case of a stable closed channel bound state, $\gamma=0$ (thick solid line, for $\Delta<\Delta_0$: $-\!\cdot\!-$ line).
$E_\mathrm{d}$ is shown as a function of the detuning $\Delta_0(I)-\Delta$ from the resonance, for a light intensity $I=1\,$kWcm$^{-2}$.
For large detunings the bound state is similar to the last bound state of the background potential (thin solid line).
The inset shows $E_\mathrm{d}(\Delta_0-\Delta)$ close to resonance, where it is approximated by $E_\mathrm{d}=-\hbar^2/(ma^2)$ ($-\!\cdot\!\cdot\,-$ line) or $E_\mathrm{d}=-\hbar^2/(m[a-\bar{a}]^2)$ (dashed line), see Sect.~\ref{sec:bgscatt}.
Approaching resonance ($\Delta\to\Delta_0(I)$), the scattering length diverges and $E_\mathrm{d}\to0$.
}
\label{fig:stableStateSpectrumHom}
\end{figure}

Suppose, for a moment, that $\gamma=0$.
In this case Eq.~(\ref{eq:EnergyDressed}) recovers relation (\ref{eq:shiftitoME}) when setting $E_\mathrm{d}=0$ and taking into account that $\Delta$ is tuned to resonance, $\Delta=\Delta_0(I)$.
This formally corresponds to the situation at a Feshbach resonance and illustrates that the scattering length diverges where the dressed state energy vanishes. 
The adiabatic energy level structure close to the scattering resonance, where the closed state crosses the background channel threshold, is shown in Fig.~\ref{fig:stableStateSpectrumHom}.
The graph shows the dependence of the then real dressed state energy $E_\mathrm{d}$ on the detuning $\Delta$ of the laser from the frequency of the transition.

We calculated this dependence using Eq.~(\ref{eq:EnergyDressed}) for photoassociation of $^{23}$Na atoms described in Sect.~\ref{sec:sodium}, but with $\gamma$ set to zero.
Technically this is achieved using the minimal model Hamiltonian which is described in Appendix \ref{app:seppot} and fixed by the background scattering length $a_\mathrm{bg}=54.9\,a_\mathrm{Bohr}$ ($a_{\rm Bohr}=0.052918\,$nm) \cite{Mies00}, the binding energy $E_{-1}/h=-318\,$MHz \cite{Elbs99}, the width $\partial\delta(I)/\partial I=38.4\,$MHz$h/($kW cm$^{-2})$, the shift $\partial\Delta_0(I)/\partial I=-164\,$MHz$h/($kW cm$^{-2})$ \cite{McKenzie02}, and the decay width $\gamma=0$. 

When approaching the resonance from the side where the scattering length is positive the bound state energy $E_\mathrm{d}$ is negative and, close to resonance, universally behaves as in Eq.~(\ref{eq:Ebgbofabar}), with $a_\mathrm{bg}$ replaced by the near resonant scattering length, Eq.~(\ref{eq:aofIandomega}) \cite{GKGJT04}.
\begin{figure}[tb]
\begin{center}
\includegraphics[width=0.45\textwidth]{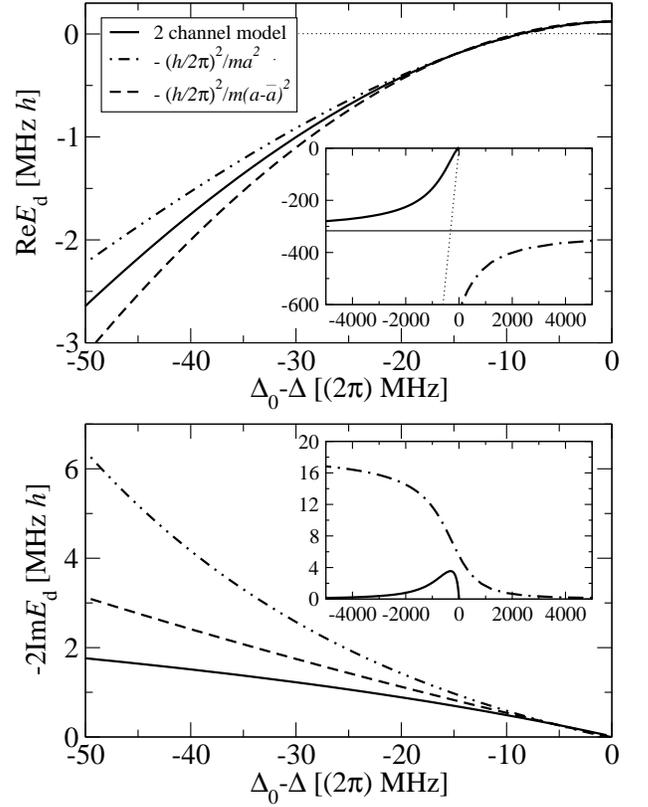}
\end{center}
\vspace*{-3ex}
\caption{
Energies $E_\mathrm{d}$ of the discrete dressed states closest to threshold, as a function of $\Delta_0(I)-\Delta$.
The graph shows the same levels as Fig.~\ref{fig:stableStateSpectrumHom}, but for a light intensity of $I=10\,$kWcm$^{-2}$ and a decay width $\gamma=(2\pi)\,18.5\,$MHz which renders the level energies complex.
The dotted and dashed lines show the analytic approximations to the near resonant behaviour as in Fig.~\ref{fig:stableStateSpectrumHom}.
Note that the real part of $E_\mathrm{d}$ may become positive, i.e.~intrude the continuum of scattering levels with real energies $E_\mathbf{p}\ge0$. 
}
\label{fig:dressedStateEnHom}
\end{figure}
This dependence of the dressed state energy on $\Delta$ is modified when the closed channel state has a nonzero width $\gamma=(2\pi)\,18.5\,$MHz.
An example of such a modified behaviour is shown in Fig.~\ref{fig:dressedStateEnHom}, for a light intensity $I=10\,$kW/cm$^2$.
In general, $E_\mathrm{d}$ is complex, and at the resonance, $\Delta=\Delta_0(I)$, neither its real nor its imaginary part vanish.
The real part rather crosses the threshold with a non-zero slope and becomes positive at $\Delta_0-\Delta\simeq -(2\pi)\,18.73\,$MHz.
At $\Delta_0-\Delta\simeq (2\pi)\,0.12\,$MHz the imaginary part vanishes and the dressed state becomes degenerate with a scattering state with non-zero energy.

We may estimate this energy by considering the analytic behaviour of $E_\mathrm{d}$ near resonance.
As in the case of a Feshbach resonance, where the binding energy approaches $E_\mathrm{b}(B)=-\hbar^2/m[a(B)]^2$ when $a(B)\rightarrow+\infty$ \cite{GKGJT04}, the same  universal dependence on $a$ applies to PA if $|a|$ is sufficiently large.
In Fig.~\ref{fig:dressedStateEnHom} we compare the real and imaginary parts of the energy given by $E_\mathrm{d}(I,\Delta)=-\hbar^2/m[a(I,\Delta)]^2$ and  $E_\mathrm{d}(I,\Delta)=-\hbar^2/m[a(I,\Delta)-\bar{a}]^2$, where $a(I,\Delta)$ is determined from Eq.~(\ref{eq:aofIandomega}), with the energy derived from Eq.~(\ref{eq:EnergyDressed}).
When crossing the resonance from the negative detuning side the dressed state is therefore transferred smoothly into a scattering state with energy $E\simeq\hbar^2\gamma^2/(4ma_\mathrm{bg}^2\delta^2)$.
This estimate is valid if $\delta\gg\gamma/4$.
For $I=10\,$kW/cm$^2$ one finds $E=$Re$E_\mathrm{d,max}/h=0.12\,$MHz.
We shall study this intrusion of the dressed state into the positive imaginary plane in more detail in the following subsection.

\subsubsection{Adiabatic energy levels in a tight trap}
\label{sec:adiablevelsTrap}
\noindent
In Sect.~\ref{sec:adiablevelsHom} we have shown that the real part of the energy of the near threshold discrete dressed state may become positive for laser detunings near resonance.
Furthermore, we have seen, that at the largest possible real part the imaginary part vanishes such that the state becomes degenerate with a state of the continuum of stable scattering states.
In the following we shall illustrate this behaviour by considering atom pairs that are confined in their motion by a tight spherical harmonic potential $v_\mathrm{trap}(r)=m\omega_\mathrm{ho}^2r^2/4$.

\begin{figure}[tb]
\begin{center}
\includegraphics[width=0.45\textwidth]{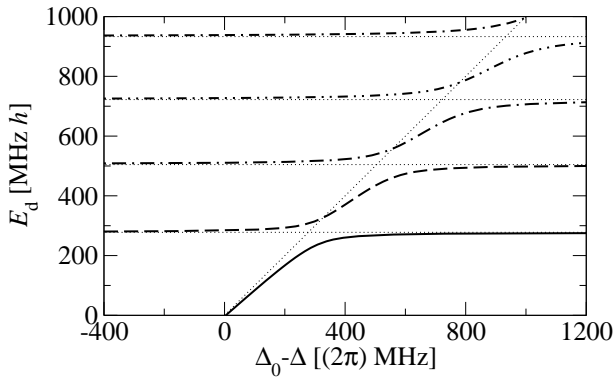}
\end{center}
\vspace*{-3ex}
\caption{
Lowest positive energy adiabatic levels of an interacting atom pair near the photoassociation resonance, for the hypothetical case of a stable closed channel bound state $\gamma=0$, as a function of the laser detuning $\Delta$. The light intensity is $I=1\,$kWcm$^{-2}$, implying a shift $\Delta_0=(2\pi)\,164\,$MHz.
The atoms are asymptotically confined in a spherical harmonic trap with oscillator frequency $\omega_\mathrm{tr}=(2\pi)\,100\,$MHz.
}
\label{fig:stableStateSpectrumTrap}
\end{figure}

For the first we consider again a stable molecular state, $\gamma=0$, such that the Hamiltonian is hermitian.
The adiabatic levels, i.e.~the dependence of the energies on the detuning $\Delta$ is generally characterized by avoided crossings of the diabatic open and closed channel states as shown in Fig.~\ref{fig:stableStateSpectrumTrap} (cf.~also, e.g., Ref.~\cite{Julienne03}).
The energies in this graph were computed for the same parameters as in Fig.~\ref{fig:stableStateSpectrumHom}, with the addition of a confining harmonic trap with oscillator frequency $\omega_\mathrm{tr}=(2\pi)\,100\,$MHz.
In Appendix \ref{app:seppotTrap} we provide the minimal Hamiltonian which enables us to describe the scattering properties in the presence of the trapping potential.

Fig.~\ref{fig:stableStateSpectrumTrap} shows that the lowest positive energy `scattering' state is raised in energy to avoid a degeneracy with the bound state approaching from below.
Similarly, the states with higher positive energy shift upwards such that the adiabatic upward ramp of the closed channel state manifests itself as an upward propagating `wavelet' of higher energy level density \cite{Julienne03}.

This level structure is modified when the closed channel state has a nonzero width $\gamma$.
As an illustration consider the non-hermitian two-level Hamiltonian
\begin{align}
  \label{eq:twolevelHam}
  H_\mathrm{2level}
  &=   
  \left(\begin{array}{cc}
   \epsilon_\mu
  &w \\
   w
  &\epsilon_\nu-i\gamma/2
  \end{array}\right),
\end{align}
with all parameters assumed to be real and positive.
At an avoided crossing of the eigenenergies of $H_\mathrm{2level}$, where $\epsilon_\mu=\epsilon_\nu$, the splitting of the levels is real if $\gamma\ge4w$ and imaginary otherwise.
Therefore it depends on the strength $w$ of the coupling whether the real parts of the energies avoid crossing.
For sufficiently weak coupling the levels cross, but the eigenenergies are both complex, with different decay widths $(\gamma/2)[1\pm\sqrt{1-16(w/\gamma)^2}]$.
\begin{figure}[tb]
\begin{center}
\includegraphics[width=0.45\textwidth]{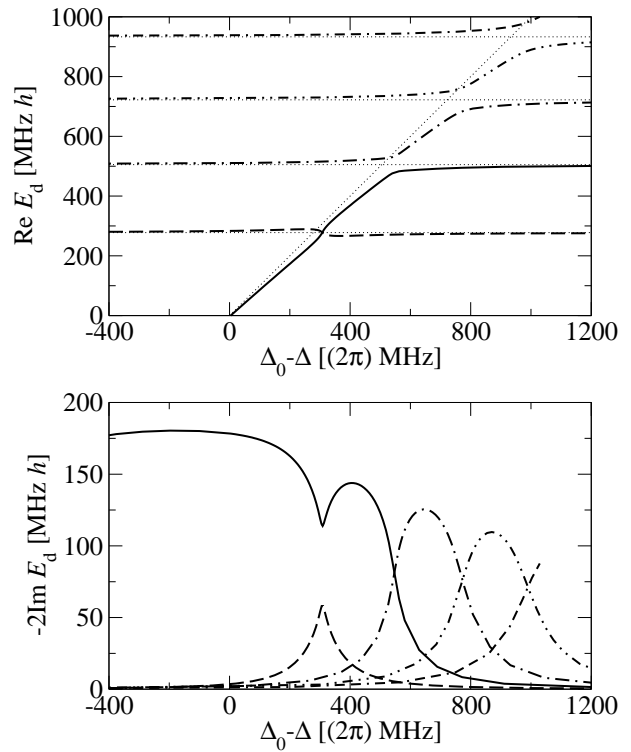}
\end{center}
\vspace*{-3ex}
\caption{
Adiabatic levels $E_\mathrm{d}(\Delta)$ of the positive energy dressed states, as a function of the laser detuning $\Delta$, for a light intensity $I=1\,$kWcm$^{-2}$, for the same trapping frequency as in Fig.~\ref{fig:stableStateSpectrumTrap}.
The width of the closed channel vibrational state was deliberately chosen on the order of the trapping frequency: $\gamma=(2\pi)\,200\,$MHz.
While the higher levels show avoided crossings as in the stable case in Fig.~\ref{fig:stableStateSpectrumTrap}, the lowest trap level avoids crossing the rising dressed state only in the imaginary part.
In the trap the scattering states become unstable in the close vicinity of the avoided crossings.
}
\label{fig:dressedStateSpectrumTrap}
\end{figure}

As a result of this the real part of the dressed state energy may, in the homogeneous case, intrude the positive energy continuum as shown in Fig.~\ref{fig:dressedStateEnHom}.
Fig.~\ref{fig:dressedStateSpectrumTrap} shows the adiabatic dependence of the real and imaginary parts of the dressed state spectrum in a $100\,$MHz trap for a hypothetical decay width $\gamma=(2\pi)\,200\,$MHz.
This illustrates in which way the dressed state `intrudes' the spectrum of `scattering' states with real and positive energy.
Whereas, at lower energies, the coupling is sufficiently weak to allow for a crossing of the real parts, scattering states with higher energy couple stronger and therefore avoid to be crossed.
The scattering states have, by virtue of their resonant coupling to the closed channel state close to the curve crossings, a finite lifetime.
However, the range of detuning over which the scattering states become unstable, is of the order of the distance of the harmonic oscillator levels, such that the scattering states are entirely real in energy in the homogeneous case considered in Fig.~\ref{fig:dressedStateEnHom}.

We would finally like to note that in Ref.~\cite{Deb03} the Franck-Condon overlaps between the different low energy trap states and the closed channel bound levels have been studied in detail.
The results corroborate our findings that the coupling strengths $w$ vary for the different level crossings as seen in Fig.~\ref{fig:dressedStateSpectrumTrap}.

\subsubsection{Two channel decomposition}
\label{sec:bgChPop}
A decisive aspect for the understanding of the two body physics close to a scattering resonance is the composition of the energy eigenstates in terms of their background and closed channel parts as given by Eq.~(\ref{eq:DressedState}).
Whereas the closed channel component has the functional form of the resonantly coupled state $|\phi_\nu\rangle$, which we have normalized to unity, the background channel component depends on $W$ and the energy $E_\mathrm{d}$ and therefore varies with $\Delta$ and $I$.
\begin{figure}[tb]
\begin{center}
\includegraphics[width=0.45\textwidth]{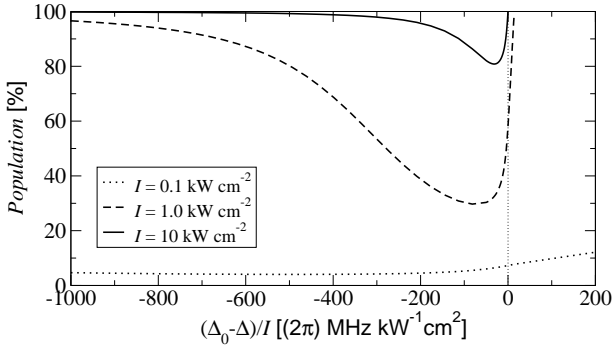}
\end{center}
\vspace*{-3ex}
\caption{
Fractional population $({\cal N}_{\rm d}^2-1)/{\cal N}_{\rm d}^2$ of the open channel component of the dressed state with energy Re$E_\mathrm{d}$ closest to resonance. 
The population is shown as a function of the detuning divided by laser intensity $(\Delta_0-\Delta)/I$, for $I=\,$0.1$\,$kW/cm$^2$ (dotted line), $I=\,$1.0$\,$kW/cm$^2$ (dashed line), and $I=\,$10.0$\,$kW/cm$^2$ (solid line).
Far away from resonance the dressed state is equivalent to the last bound state of the background potential such that $({\cal N}_{\rm d}^2-1)/{\cal N}_{\rm d}^2$ approaches 100\%.
Closer to resonance the closed channel state dominates the dressed state for not too large intensities $I$.
The population, however, reaches again 100\% at the resonance, where the dressed state goes over in a stable scattering state (cf.~Sect.~\ref{sec:adiablevelsTrap}).
}
\label{fig:bgChPopofDelta}
\end{figure}

In Figure \ref{fig:bgChPopofDelta} we show the population 
\begin{align}
  \label{eq:bgChPop}
  \langle\phi^\mathrm{bg}_\mathrm{d}|\phi^\mathrm{bg}_\mathrm{d}\rangle
  &= \frac{{\cal N}_\mathrm{d}^2-1}{{\cal N}_\mathrm{d}^2}
\end{align}
of the background channel component of the dressed state as a function of the intensity-renormalized detuning $(\Delta_0-\Delta)/I$, for $I=0.1\,$kW/cm$^2$,  $I=1.0\,$kW/cm$^2$, and $I=10\,$kW/cm$^2$.
${\cal N}_\mathrm{d}$ is given by Eq.~(\ref{eq:NormDressed}).
Compare this to the corresponding dependence of the dressed state energy and width shown in Fig.~\ref{fig:dressedStateEnHom}.
Far from resonance, $\Delta\gg\Delta_0(I)$, where the dressed state goes over into the uppermost background channel bound state, the population is near 100\%.
In the intermediate region, the population decreases significantly and the dressed state is dominated by its closed channel state.

Closer to resonance, however, the weight of the background channel component rises again abruptly.
On resonance, $\Delta=\Delta_0(I)$ (vertical dotted line), the population is near 50\%, and rises on the positive energy side to 100\% where the detuning $\Delta$ is such that the imaginary part of the dressed state energy vanishes.
For $I=0.1\,$kW/cm$^2$ (dotted curve) we have not extended the plot to the detuning $(\Delta_0-\Delta)/I=(2\pi)\,1.12\cdot10^4\,$MHz kW$^{-1}$ cm$^2$ where the population has risen to 100\%.
\begin{figure}[tb]
\begin{center}
\includegraphics[width=0.45\textwidth]{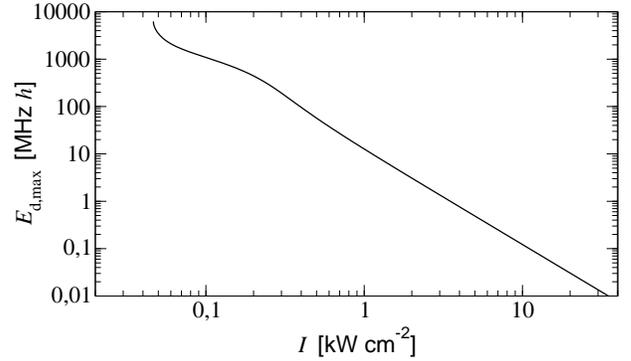}
\end{center}
\vspace*{-3ex}
\caption{
``Intrusion'' of the uppermost dressed state into the positive energy scattering continuum. Shown is the maximum energy $E_\mathrm{d,max}=\mathrm{Re}(E_\mathrm{d}(\Delta,I))$ as a function of $I$, for an appropriately chosen detuning $\Delta$.
Below $I=0.047\,$kW/cm$^2$ the intrusion is infinite.
For $I>0.5\,$kW/cm$^2$ the maximum intrusion decreases to zero with a negative power of $I$.
}
\label{fig:EdmaxofI}
\end{figure}
This examples shows that the intrusion of the dressed channel state into the positive energy continuum as described in the previous section rises with decreasing intensity.
In fact, we found that at $I=0.047\,$kW/cm$^2$ the intrusion reaches $6.2\,$GHz, and below this intensity it is unlimited, meaning that for any positive detuning $(\Delta_0-\Delta)/I$, a discrete dressed state with an energy with positive real part exists, see Fig.~\ref{fig:EdmaxofI}.

For large positive $(\Delta_0-\Delta)/I$ the imaginary part then also approaches $\gamma/2$. 
This reflects the complete decoupling of the dressed state from the scattering states as one would expect from considering the corresponding Franck-Condon overlaps.

\begin{figure}[tb]
\begin{center}
\includegraphics[width=0.45\textwidth]{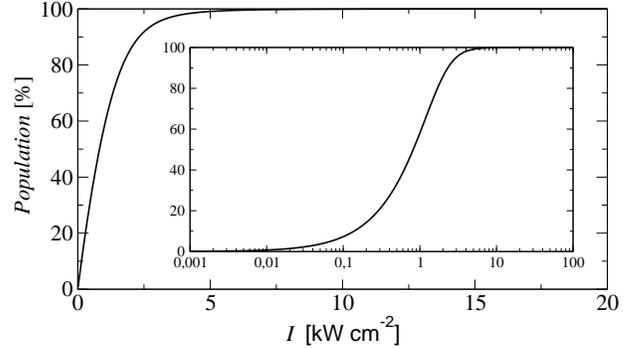}
\end{center}
\vspace*{-3ex}
\caption{
Population $({\cal N}_{\rm d}^2-1)/{\cal N}_{\rm d}^2$ of the open channel component of the dressed state wave function as a function of the intensity $I$ of the laser which is tuned to the corresponding ($I$-dependent) resonance frequency, $\Delta=\Delta_0(I)$.
For low intensities this population is small implying the dressed state to be similar to the closed channel bound state.
In contrast, strong laser coupling prohibits the dressed state to further intrude into the scattering continuum such that it rather resembles the zero energy scattering state whose closed channel contribution is negligible. 
}
\label{fig:bgChPopofI}
\end{figure}
\begin{figure}[tb]
\begin{center}
\includegraphics[width=0.45\textwidth]{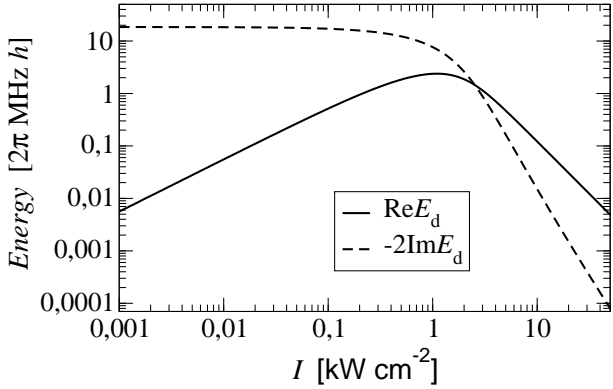}
\end{center}
\vspace*{-3ex}
\caption{
Real and imaginary parts of the energy $E_\mathrm{d}(\Delta=\Delta_0(I))$ of the near threshold dressed state on resonance as a function of the laser intensity $I$.
As anticipated from Figs.~\ref{fig:dressedStateSpectrumTrap} and \ref{fig:bgChPopofI} the dressed state is pushed out of the scattering continuum for strong couplings, $I\gg1\,$kW/cm$^2$.
}
\label{fig:EdonResofI}
\end{figure}

In the experiment by McKenzie {\it et al.}, Ref.~\cite{McKenzie02}, the detuning of the laser was fixed whereas the intensity $I$ was varied in time.
Let us therefore have a look at the dependence of the background channel population and the dressed state energy for the case of different intensities $I$, while $\Delta=\Delta_0(I)$ chosen on resonance.
Fig.~\ref{fig:bgChPopofI} shows the dependence of the on-resonance open channel population on $I$, as determined from Eq.~(\ref{eq:bgChPop}).
At low energies the dressed state resembles the closed channel bound state $|\phi_\nu\rangle$.
The higher the intensity $I$, however, the less significant is the closed channel part of the dressed state.
This may have been anticipated from the schematic dressed state spectrum in a trap shown in Fig.~\ref{fig:dressedStateSpectrumTrap}.
For sufficiently strong coupling, i.e.~large $I$, the energy $\mathrm{Re}E_\mathrm{d}$ avoids crossing with the lowest scattering state which implies that the dressed state goes over into this scattering level as soon as it crosses threshold at $\Delta=\Delta_0$. 

In Fig.~\ref{fig:EdonResofI} we explicitly show the dependence of the real and imaginary parts of the on resonance energy $E_\mathrm{d}(\Delta=\Delta_0(I))$ on the intensity $I$. 
The intrusion of the on resonance dressed state energy into the positive energy continuum rises with $I$ before the dressed state is pushed out again at the highest intensities, where the situation resembles the case of a stable ($\gamma=0$) closed channel state, i.e.~$E_\mathrm{d}\rightarrow0$.

\subsubsection{Universal shape}
\label{sec:bgChShape}
It was predicted in Ref.~\cite{KGJB03} that close to a Feshbach resonance the uppermost bound state is dominated by its background channel component which is spatially extended over a very large range of the order of the scattering length $a$.
The extremely weekly bound molecules formed on the side with positive $a$ have binding lengths of unprecedented size, ranging up to several thousand $a_\mathrm{Bohr}$.
The production of such molecules with ultralong binding length through photoassociation of a trapped atom pair has been studied in Ref.~\cite{Deb03}.
Generalizing the arguments in Ref.~\cite{GKGJT04} to the case where $\gamma>0$ one finds that on interatomic distances $r$ large compared to the van der Waals length the background channel component depends on $r$ as
\begin{align}
  \label{eq:phibofr}
  \phi^\mathrm{bg}_\mathrm{d}(r)
  &= \sqrt{\mathrm{Re}\left(\frac{1}{2\pi a(I,\Delta)}\right)}
     \frac{e^{-r/a(I,\Delta)}}{r}.  
\end{align}
From this universal asymptotic behaviour a remarkable feature of PA to an unstable molecular state may be inferred which contrasts the case of a Feshbach resonance where spontaneous decay is suppressed.
Eq.~(\ref{eq:aofIandomega}) yields Re$[1/a(I,\Delta_0(I))]=a_\mathrm{bg}/|a(I,\Delta_0(I))|^2$, i.e., according to Eq.~(\ref{eq:phibofr}), the spatial extension $\langle\phi_\mathrm{d}^\mathrm{bg}|r|\phi_\mathrm{d}^\mathrm{bg}\rangle=1/(2\mathrm{Re}[1/a(I,\Delta)])$ of the resonant dressed state is \emph{enhanced by a factor of} $|a|/a_\mathrm{bg}\simeq2\delta(I)/\gamma$.
Consequently, for $I=\,$10$\,$kW cm$^{-2}$ the dressed state has an extension which is approximately two orders of magnitude larger than for $I=\,$1$\,$kW cm$^{-2}$. 

\begin{figure}[tb]
\begin{center}
\includegraphics[width=0.45\textwidth]{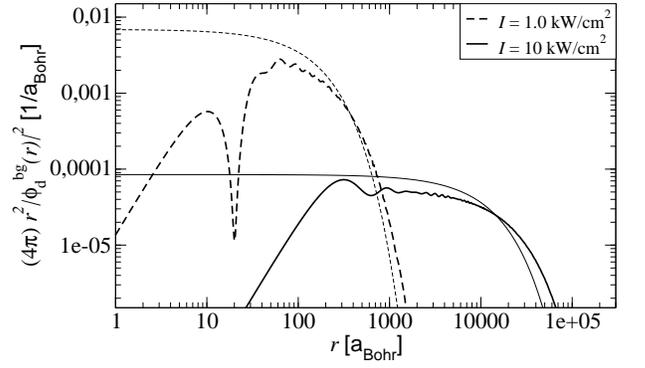}
\end{center}
\vspace*{-3ex}
\caption{
Open channel component of the radial dressed state density for $I=\,$1.0$\,$kW/cm$^2$ (dashed lines) and $I=\,$10$\,$kW/cm$^2$ (solid lines), on the PA resonance, $\Delta=\Delta_0(I)$.
$r$ and the density are given on logarithmic scales.
The thin lines show the respective long range approximations according to Eq.~(\ref{eq:phibofr}).
The spatial extension of the on resonance wave function rises approximately $\propto I^2$.
}
\label{fig:phidbgofr}
\end{figure}
Fig.~\ref{fig:phidbgofr} shows the radial probability distribution $|\phi^\mathrm{bg}_\mathrm{d}(r)|^2$ associated with background component of the highest vibrational dressed state. 
The wave function was determined using the minimal Hamiltonian introduced in Appendix \ref{app:seppot}, for intensities $I=\,$1.0$\,$kW cm$^{-2}$ and $I=\,$10$\,$kW cm$^{-2}$ of the laser, with its frequency tuned on resonance, $\Delta=\Delta_0(I)$.
For comparison, also the asymptotic form of the density as approximated by the analytic expression in Eq.~(\ref{eq:phibofr}) is given.
This has to be compared to the radial extension of the wave functions away, though close to resonance, which is shown in Fig.~4 of Ref.~\cite{G0104} which, at high intensities does not show the extra amplification factor.

\section{Many body dynamics of a Bose-Einstein condensate under single color photoassociation}
\label{sec:manybodytheory}
In the vicinity of a photoassociative resonance the low energy scattering properties are modified due to the presence of a dressed state near threshold.
This has been discussed in detail in Sect.~\ref{sec:2bodytheory}.
For sufficiently high light intensities the modulus of the scattering length is significantly increased near resonance, with in general strong consequences on the dynamics of a Bose-Einstein condensed many body ensemble.

In this section we shall investigate the many body dynamics of a trapped Bose-Einstein condensate which is exposed to a laser pulse tuned to a fixed frequency near resonance.
We will focus on possible limitations of the maximum rate at which condensate atoms may be associated to diatomic molecules.
These limits concern, on the one hand, the rates at which atoms are lost from the condensate, and, on the other hand, the maximum efficiency with which these atoms may be associated to ground state molecules.
We will show in detail how differences may arise between these limits.

We will first adapt the dynamical many body theory introduced in Refs.~\cite{KB02,KGB03,KGG03,GKGJT04} to the case of single color photoassociation to an unstable closed channel bound state. Central results have been presented in Ref.~\cite{G0104}. 

We shall study the dynamical evolution of the condensate as well as of the total number of atoms within the subspace defined by their relative motional state being of the form described in Sect.~\ref{sec:2chdressed}.
The many body theory allows to systematically treat the condensate mean field and the higher order correlation functions.
To determine the mean field evolution we will take into account the dynamics of the pair correlation function.
The results of Sect.~\ref{sec:2bodytheory} allow us to rigorously treat the two body physics within the many body formalism.
To the order of approximation chosen here the many body theory includes processes where colliding pairs of condensate atoms are transferred to the two body dressed state as well as into scattering states with higher relative energy.
Moreover, atom pairs in the dressed state may be lost from the overall system through spontaneous decay caused by the nonzero width $\gamma$ of the closed channel state.

\subsection{Microscopic Dynamical Many Body Theory}
\label{sec:mqda}
The many body theory we will adapt for our purposes has been introduced in Refs.~\cite{KB02,KGB03}.
Its applications presented so far include the production of correlated atom pairs in four wave mixing of condensates \cite{KB02}, the second order correction to the condensate ground state energy including the effects of three-body collisions \cite{Koehler02}, the atom-molecule coherence in a Ramsey type interferometer \cite{KGB03}, and the condensate and molecular dynamics in Feshbach crossing experiments \cite{KGG03,GKGJT04}.
A detailed discussion of its application to Fesh\-bach resonances within a two-channel picture and of its relation to other extended mean field descriptions \cite{Kokkelmans02,Mackie02,Yurovsky03} of the near resonant many body physics has been presented in \cite{GKGJT04}.

The theory is based on a resummation of the dynamical equations for the single time many body correlation functions in terms of cumulants which may be characterized as connected Green's functions \cite{Fricke96,Koehler02,KGB03}.
As the general technique have been described in detail before \cite{KB02,KGB03,GKGJT04}, we shall concentrate on those aspects specific for its application to single color photoassociation and required for the subsequent discussion.

\subsubsection{Two channel dynamic equations}
\label{sec:2chdyneq}
The general multichannel many body equations \cite{GKGJT04} summarized in Appendix~\ref{sec:multichannel_mb} allow to describe a variety of different photoassociation schemes in a gas consisting of an arbitrary number of different bosonic species.
In the following we will specify the equations to PA with a single laser the two body description of which has been given in Sect.~\ref{sec:2bodytheory}.
We may thus restrict the many body formalism to the background scattering channel coupled via the laser to a closed channel molecular state as described in Sect.~\ref{sec:sodium}.

In accordance with this picture an atomic condensate mean field only exists for ground state atoms.
We denote it by $\Psi(\mathbf{x},t)$.
Since the laser transition is far detuned from the excimer potential asymptote, the channel is closed, and a macroscopic population of the excited state may be neglected.
To take into account the two body correlations which are, in the many body system, a measure for the population of the binary dressed states \cite{KGB03}, we can restrict the matrix of pair functions (cf.~App.~\ref{sec:multichannel_mb}) to its elements corresponding to the background channel spectrum, $\Phi_{\rm bg}(\mathbf{x},\mathbf{y},t)$, and the closed channel state, $\Phi_{\rm cl}(\mathbf{x},\mathbf{y},t)$.

The mean field for ground state atoms and the pair functions then obey the non linear dynamic equations (cf.~Eqs.~(\ref{eq:DynEqPsik}), (\ref{eq:DynEqPhikl})):
\begin{widetext}
\begin{align}
 \label{eq:meanfield}
  \nonumber\\[-1cm]
  i\hbar\frac{\partial}{\partial t}\Psi(\mathbf{x},t)
  = H_{\rm g}^\mathrm{1B}(\mathbf{x})
  \Psi(\mathbf{x},t)
  +\!\!\int\! d\mathbf{y}\Psi^*\!(\mathbf{y},t)\Big(
  V_{\rm bg}(|\mathbf{x}\!-\!\mathbf{y}|)
  \Big[
    \Phi_{\rm bg}(\mathbf{x},&\mathbf{y},t)+
    \Psi(\mathbf{x},t)\Psi(\mathbf{y},t)
    \Big]
    \!+\!\mbox{$\frac{\hbar}{2}$}\Omega_0(t)\,P_\nu
    \Phi_{\rm cl}(\mathbf{x},\mathbf{y},t)\Big),
  \\
 \label{eq:pairfunction}
  i\hbar\frac{\partial}{\partial t}
  \left(\begin{array}{c}
        \Phi_{\rm bg}(\mathbf{x},\mathbf{y},t)\\
	\Phi_{\rm cl}(\mathbf{x},\mathbf{y},t)
  \end{array}\right)
  = H^\mathrm{2B}(\mathbf{x},\mathbf{y},t)
  \left(\begin{array}{c}
        \Phi_{\rm bg}(\mathbf{x},\mathbf{y},t)\\
	\Phi_{\rm cl}(\mathbf{x},\mathbf{y},t)
  \end{array}\right)+
  &\left(\begin{array}{c}
        V_{\rm bg}(|\mathbf{x}-\mathbf{y}|)\\
	P_\nu\,\frac{\hbar}{2}\Omega_0(t)
  \end{array}\right)
  \Psi(\mathbf{x},t)\Psi(\mathbf{y},t).
\end{align}
\end{widetext}
Here $H^\mathrm{1B}_{\rm g}(\mathbf{x})=-\hbar^2\nabla^2/2m+V_\mathrm{trap}(\mathbf{x})$ is the Hamiltonian of a single atom in the electronic ground state.
Furthermore, $H^\mathrm{2B}(\mathbf{x},\mathbf{y},t)$ is the general two-channel two-body Hamiltonian of interacting atoms which in the interaction picture and rotating wave approximation has the form (cf.~Eqs.~(\ref{eq:H2B0nu}), (\ref{eq:V2Bnu})): 
%
\begin{align}
 \label{eq:H2B0I}
  H_\mathrm{2B}(t)
  &= \left(\begin{array}{cc}
      T_\mathrm{kin} & 
      0 \\
      0  & 
      -\hbar(\Delta(t)+i\gamma/2)\,P_\nu
      \end{array}\right)
      + V_\mathrm{2B}(t)
 \\
 \label{eq:V2BI}
 V_\mathrm{2B}(t)
  &= \left(\begin{array}{cc}
      V_{\rm bg} & 
      \frac{\hbar}{2}\Omega_0(t)\,P_\nu \\
      P_\nu\,\frac{\hbar}{2}\Omega_0(t)  & 
      0
      \end{array}\right).
\end{align}
%
$H_\mathrm{2B}$ may, in general, depend on time through the detuning $\Delta(t)=\omega(t)-\omega_\nu$ and the intensity of the light $I(t)$ which enters the Rabi frequency $\Omega_0(t)$.

The transformation to the interaction picture replaces the Hamiltonian in Eqs.~(\ref{eq:H2B0nu}), (\ref{eq:V2Bnu}) by the Hamiltonian in Eq.~(\ref{eq:H2B0I}).
Thereby, it reduces the coupling $W(\mathbf{R},t)=\Omega_0(t)\cos(\mathbf{k}\mathbf{R}-\omega t)$, with $\mathbf{R}=(\mathbf{x}+\mathbf{y})/2$, to the envelope $\Omega_0(t)/2$ and introduces a transformed closed channel pair function also denoted as $\Phi_\mathrm{cl}$.
As we chose the zero of the energy to coincide with the background potential asymptote, the functions $\Psi$ and $\Phi_\mathrm{bg}$ are invariant under the transformation to the interaction picture.
Physically the transformation of $\Phi_\mathrm{cl}$ keeps track of the fact that the atom pair in the closed channel bound state takes up energy and momentum from the photon.
Note also that we may neglect the phase difference of the light between positions $\mathbf{x}$ and $\mathbf{y}$ since the coupling $WP_\nu$ acts only over distances of the order of the binding length of the closed channel bound state which is much shorter than the wave length of the laser. 

For solving the system of dynamic equations (\ref{eq:meanfield}) and (\ref{eq:pairfunction}) numerically we found it convenient to formally solve the inhomogeneous linear equation (\ref{eq:pairfunction}) in terms of the two body time evolution operator $U_\mathrm{2B}(t,t_0)={\cal T}\exp\{-(i/\hbar)\int_{t_0}^tdt'H_\mathrm{2B}(t')\}$ and to insert this solution into the non-linear equation (\ref{eq:meanfield}).
The dynamics of the atomic mean field is then determined by the single non-Markovian non-linear equation:
\begin{widetext}
\begin{align}
  \label{eq:NMNLSE}
  i\hbar\frac{\partial}{\partial t}\Psi(\mathbf{x},t)
  &= 
  H^\mathrm{1B}_{\rm g}(\mathbf{x})
  \Psi(\mathbf{x},t) 
  - \Psi^*(\mathbf{x},t)\left[
  \int_{t_0}^\infty d\tau
  \Psi^2(\mathbf{x},\tau)\frac{\partial}{\partial \tau}h(t,\tau)
  +(2\pi\hbar)^{3/2}
  \langle0,\mathrm{bg}|V_\mathrm{2B}(t)U_\mathrm{2B}(t,t_0)|\Phi(t_0)\rangle
  \right],
\end{align}
\end{widetext}
where  
$h(t,\tau)=(2\pi\hbar)^3\langle0,\mathrm{bg}|V_\mathrm{2B}(t)
U_\mathrm{2B}(t,\tau)|0,\mathrm{bg}\rangle\,\theta(t-\tau)$. 
$V_\mathrm{2B}$ is given by Eq.~(\ref{eq:V2BI}), and $\theta(t-\tau)$ is the step function that evaluates to zero except for $t>\tau$.
In deriving Eq.~(\ref{eq:NMNLSE}) we have taken into account that the spatial range of the background and closed channel potentials is much smaller than the scale of the variation of the cumulants.
This allows us to eliminate the spatial integrations and leads to the zero momentum plane wave state $|0,\mathrm{bg}\rangle$ of the relative motion of two atoms in the background channel \cite{KGB03,GKGJT04}.
In the second term in the square brackets $|\Phi(t_0)\rangle$ represents pair correlations at the initial time $t_0$ which will be discussed in following section.

\subsubsection{Initial correlations}
\label{sec:inicor}
The solution of the mean field dynamic equation (\ref{eq:NMNLSE}) depends on the initial condensate wave function $\Psi(\mathbf{x},t_0)$ as well as on the initial pair correlations $|\Phi(t_0)\rangle$. These are denoted as a `ket' in the sense of $\langle\mathbf{x},\mathbf{x'}|\Phi(t)\rangle=|\mathrm{bg}\rangle\Phi_\mathrm{bg}(\mathbf{x},\mathbf{x'},t)+|\mathrm{cl}\rangle\Phi_\mathrm{cl}(\mathbf{x},\mathbf{x'},t)$.
In general these initial conditions are given by the experimental setup.

For a weakly interacting dilute Bose-Einstein condensed gas the non-Markovian mean field equation should recover, in the absence of the photoassociation laser, the stationary solution of the Gross-Pitaevskii equation (GPE).
One may fulfill this requirement by an appropriate choice of the initial pair correlations $|\Phi_0\rangle$.
Neglecting the decoupled closed channel component $\Phi_\mathrm{cl}$, we choose the \emph{ansatz}:
\begin{widetext}
\begin{align}
  \nonumber
  \Phi_\mathrm{bg}(\mathbf{R},\mathbf{r},t_0) =
  \frac{1}{2}(2\pi\hbar)^3n_0(\mathbf{R})
  &\Bigg[
  \int_{t_0}^\infty d\tau\, e^{-i2\mu_0(\tau-t_0)/\hbar}
  \frac{\partial}{\partial\tau}
  \langle\mathbf{r},\mathrm{bg}|U_\mathrm{2B}(t,\tau)|0,\mathrm{bg}\rangle
  \\
  &-\int^{t_0}_{-\infty} d\tau\, e^{-i2\mu_0(\tau-t_0)/\hbar}
  \frac{\partial}{\partial\tau}
  \langle\mathbf{r},\mathrm{bg}|U_\mathrm{2B}(t,\tau)|0,\mathrm{bg}\rangle
  \Bigg].
  \label{eq:inicor}
\end{align}
\end{widetext}
Here $\mathbf{R}=(\mathbf{x}+\mathbf{y})/2$ and $\mathbf{r}=\mathbf{x}-\mathbf{y}$ are the center of mass and relative coordinates, respectively. 
$n_0$ is the equilibrium condensate density, and the chemical potential $\mu_0$ is determined self consistently from the stationary non-linear Schr\"odinger equation:
\begin{align}
  \label{eq:statGPE}
  \mu_0 n_0^{1/2}(\mathbf{x})
  &= [T_\mathrm{kin}+V_\mathrm{trap}(\mathbf{x})
  +g(\mu_0)n_0(\mathbf{x})]n_0^{1/2}(\mathbf{x})
\end{align}
In this equation, the energy dependent coupling constant $g$ involves the $T$-matrix associated with the scattering in the background channel potential: $g(\mu_0)=(2\pi\hbar)^3\mathrm{Re}[\langle 0|T_\mathrm{bg}(2\mu_0+i0)|0\rangle]$ (cf., e.g., Ref.~\cite{KB02}).

In the dilute gas limit which one has if for the peak density the condition $n_\mathrm{peak}a_\mathrm{bg}^3\ll1$) holds, Eq.~(\ref{eq:statGPE}) recovers the stationary GPE when $g$ is evaluated at $\mu_0\rightarrow0$.
Note that Eq.~(\ref{eq:statGPE}), in the limit $t\rightarrow\infty$, \emph{exactly} follows from the dynamic equation (\ref{eq:NMNLSE}) if one inserts the stationary ansatz $\Psi(\mathbf{x},t)=n_0^{1/2}(\mathbf{x})\exp[-i\mu_0(t-t_0)/\hbar]$ for the mean field for $t\to\infty$.
 
We finally remark that instead of starting with the ansatz in Eq.~(\ref{eq:inicor}) one could have implemented the stationarity condition by choosing the \emph{ans\"atze}
\begin{align}
  \label{eq:statPsiAnsatz}
  \Psi(\mathbf{R},t)
  &= \sqrt{n_0(\mathbf{R})}\,e^{-i\mu_0(t-t_0)/\hbar},
  \\
  \label{eq:statPhiAnsatz}
  \Phi_\mathrm{bg}(\mathbf{R},\mathbf{r},t)
  &= n_0(\mathbf{R})\,e^{-i2\mu_0(t-t_0)/\hbar},
\end{align}
for the mean field and background channel pair function and derive the initial pair function $\Phi_\mathrm{bg}(\mathbf{R},\mathbf{r},t_0)$ from the formal solution of Eq.~(\ref{eq:pairfunction}), cf.~Ref.~\cite{KGG03}.

We underline that the stationarity of the solution for the mean field requires, in our dynamical formulation, a carefully balanced equilibrium between the condensate and the fraction of the gas described by the pair function.
When not taking into account the pair correlations in the initial state the dynamic equation (\ref{eq:NMNLSE}) yields, even in the absence of the laser coupling, a steadily shrinking condensate fraction.
This loss rather quickly exceeds the stationary depletion resulting from Eq.~(\ref{eq:statGPE}), roughly after a time $t_\mu=\hbar/|\mathrm{Im}\mu|$.
Here $\mathrm{Im}\mu=(4\pi\hbar^2a_\mathrm{bg}n/m)(8\pi na_\mathrm{bg}^3)^{1/2}=\mathrm{Im}[-a_\mathrm{bg}(1-ika_\mathrm{bg})]$ is the imaginary part of $\langle 0,\mathrm{bg}|T_{2B}(2\mu+i0)|0,\mathrm{bg}\rangle$ evaluated to leading order in the momentum $k=\sqrt{2m\mathrm{Re}\mu}/\hbar$ which corresponds to the energy of the condensate atoms.

In the photoassociation experiment \cite{McKenzie02} the peak condensate density is of the order of $4\times10^{14}\,$cm$^{-3}$, such that the gas, without the laser being switched on, is weakly interacting: $n_\mathrm{peak}a_\mathrm{bg}^3\simeq 1\cdot10^{-5}$.
However, the density is sufficiently high that the time scale $t_\mu$ is on the order of $t_\mu\simeq 1.6\,$ms, which is to be compared with the typical duration of $100\,\mu$s of the applied laser pulse.
Taking into account the initial correlations $|\Phi(t_0)\rangle$ therefore ensures that at low light intensities where the induced condensate loss is weak, Eq.~(\ref{eq:NMNLSE}) yields the correct time evolution.

\subsubsection{Total density}
\label{sec:totden}
In deriving the dynamic equation (\ref{eq:NMNLSE}) for the mean field $\Psi(\mathbf{x},t)$ we neglected, in Sections \ref{sec:multichannel_mb} and \ref{sec:2chdyneq}, the time evolution of the density $\Gamma_{\kappa\lambda}$ of non-condensed atoms, Eq.~(\ref{eq:Gammakl}).
In Refs.~\cite{KB02,KGB03} it was shown that this is justified if the thermal fraction of the gas is negligible at the initial time $t_0$.
Moreover, as mentioned in Sect.~\ref{sec:multichannel_mb}, in the leading order approach a non-condensed fraction arises from the buildup of pair correlations, and, in the background and closed channels, is given by
\begin{align}
  \label{eq:Gammabgcl}
  \left(\begin{array}{c}
        \Gamma_{\rm bg}(\mathbf{x},\mathbf{y},t)\\
	\Gamma_{\rm cl}(\mathbf{x},\mathbf{y},t)
  \end{array}\right)
  &= \int\,d\mathbf{x'}\, 
  \left(\begin{array}{c}
        \Phi_{\rm bg}  (\mathbf{x},\mathbf{x'},t)
	\Phi_{\rm bg}^*(\mathbf{y},\mathbf{x'},t)\\
	\Phi_{\rm cl}  (\mathbf{x},\mathbf{x'},t)
	\Phi_{\rm cl}^*(\mathbf{y},\mathbf{x'},t)
  \end{array}\right)
\end{align}
The total density of the gas thus is (cf.~Eq.~(\ref{eq:NtotMulCh})):
\begin{align}
  \label{eq:ntot}
  n_\mathrm{tot}(\mathbf{x},t)
  &=\Gamma_{\rm bg}(\mathbf{x},\mathbf{x},t)
   +\Gamma_{\rm cl}(\mathbf{x},\mathbf{x},t) 
   +|\Psi(\mathbf{x},t)|^2
\end{align}
Hence, atoms lost from the condensate are transferred to the non-condensate fraction in the form of pair correlations $\Phi\not=0$.
Due to the finite life time of the closed channel state they may then be lost from the overall system leading to a decrease of $N_\mathrm{tot}(t)=\int\,d\mathbf{x}\,n_\mathrm{tot}(\mathbf{x},t)$ with time $t$.
As is shown in Appendix \ref{app:optth} the time evolution of the total number of atoms  is determined by
\begin{widetext}
\begin{align}
  \label{eq:dNndt}
  \frac{\partial}{\partial t}N_\mathrm{tot}(t)
  &= -\gamma\int\,d\mathbf{R}\,\left|
     \langle\Phi(t_0)|U_\mathrm{2B}^\dagger(t,t_0)|\phi_\nu,\mathrm{cl}\rangle
     -(2\pi\hbar)^{3/2}\int_{t_0}^t d\tau 
     \Psi^2(\mathbf{R},\tau)\frac{\partial}{\partial \tau}e^{i\varphi(t)}
     \langle\tilde\phi_\nu,\mathrm{cl}|
     U_\mathrm{2B}^\dagger(t,\tau)
     |0,\mathrm{bg}\rangle\right|^2,
\end{align}
\end{widetext}
where $\varphi(t)=\int_{t_0}^t\,d\tau\,(\Delta(\tau)+i\gamma/2)$.
Note that the term including $|\Phi(t_0)\rangle$ depends on $\mathbf{R}$ according to Eq.~(\ref{eq:inicor}).
For $\gamma=0$ the total number $N_\mathrm{tot}$ would be conserved in time.

\subsection{Numerical results for photoassociation in $^{23}$Na}
\label{sec:manybodyresults}
The microscopic theory summarized in Sect.~\ref{sec:mqda} sets us in the position to study the many body dynamics of an ultracold gas under the influence of a photoassociation laser. 
Starting from a dilute Bose-Einstein condensate of a trapped gas of sodium atoms we shall now present the numerical solutions of the dynamical equations (\ref{eq:NMNLSE}) and (\ref{eq:dNndt}).
We will first consider the condensate evolution and study the effects of the non-Markovian character of the interaction term in Eq.~(\ref{eq:NMNLSE}) thereby concentrating on the achievable photoassociation loss rates.

\subsubsection{Atomic condensate fraction}
\label{sec:condevol}
Solving the non-linear non-Markovian mean field equation (\ref{eq:NMNLSE}) requires the knowledge of the time evolution operator $U_\mathrm{2B}(t,\tau)$ which contains all necessary information about the two body interactions under the influence of the laser pulse.
The fact that the equations involve only expectation values of this operator and its product with the potential matrix $V_\mathrm{2B}$ in the plane wave state $|0,\mathrm{bg}\rangle$ reduces the numerical effort significantly.
We shall use the minimal model Hamiltonian introduced in Appendix \ref{app:seppotHom} to determine the required evolution kernels as described in Appendix \ref{app:seppotTimeEvol}.

As initial state we chose a condensate of $N_\mathrm{c}(t_0)=4.0\cdot10^6$ $^{23}$Na atoms in a spherical harmonic trap with oscillator frequency $\nu_\mathrm{ho}=198\,$Hz.
This corresponds to the situation in the experiment in Ref.~\cite{McKenzie02}, with the anisotropic harmonic trap replaced by a spherically symmetric one with mean frequency $\nu_\mathrm{ho}=(\nu_x\nu_y\nu_z)^{1/3}$.
Solving the GPE gave the initial mean field $\Psi(\mathbf{R},t_0)$ describing this condensate. 
The initial background channel pair function $\Phi_\mathrm{bg}(\mathbf{R},\mathbf{r},t_0)$ is then given by Eq.~(\ref{eq:inicor}), with the asymptotic density $n_0(\mathbf{R})=|\Psi(\mathbf{R},t_0)|^2$.

\begin{figure}[tb]
\begin{center}
\includegraphics[width=0.45\textwidth]{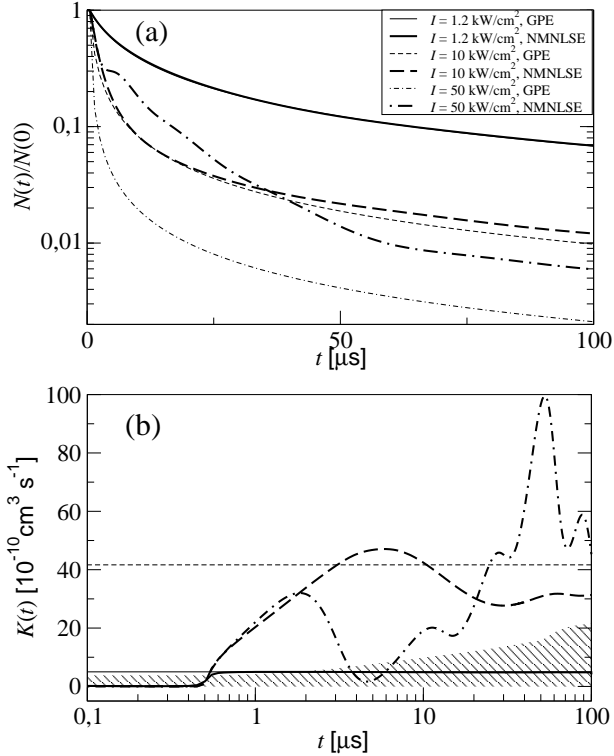}
\end{center}
\vspace*{-3ex}
\caption{
(a) Time evolution of the remaining condensate fraction of the initial atom number $N(0)\!\!=\,$4.0$\,\cdot\,$10$^{6}$ during a $100\,\mu$s pulse, with peak intensities of $1.2\,$kW/cm$^2$ (solid lines), $10\,$kW/cm$^2$ (dashed lines), and $50\,$kW/cm$^2$ (dash-dotted lines). 
The fraction $N(t)/N(0)$ is given on a logarithmic scale.
Shown are the solutions of the non-Markovian non-linear Schr\"odinger equation (NMNLSE) (\ref{eq:NMNLSE}) as thick lines and, for comparison, the evolution according to the Gross-Pitaevskii equation (GPE) as thin lines,  $n(\mathbf{x},t)=n(\mathbf{x},0)[1+K_0n(\mathbf{x},0)(t-t_{\rm rise})]^{-1}$, $t_{\rm rise}=0.5\,\mu$s, with $K_0=(8\pi\hbar/m)|\Im a(\Delta,I)|$.
(b) Time evolution of the temporally local decay `rate' $K(t)=-\dot{N}(t)/\int d\mathbf{x}\,n^2(\mathbf{x},t)$ corresponding to the cases considered in (a).
The GPE rates are indicated as horizontal lines, except for the rate for $I=50\,$kW/cm$^2$, which is $208\cdot10^{-10}\,$cm$^3/$s.
Time is shown on a logarithmic scale.
Since due to the light induced shift $\Delta_0(I)$ the resonance condition is first met at the end of the initial ramp, $t=0.5\,\mu$s, the NMNLSE rates are close to zero before.
The hatched area indicates the values of $K(t)$ permitted according to the limit found in Refs.~\protect\cite{Kostrun00,Javanainen02}, cf.~Sect.~\ref{sec:condratelimit}.
}
\label{fig:condtimedep}
\label{fig:locdecayrate}
\end{figure}
Starting at time $t_0=0$, the laser light, with fixed frequency $\omega=\mathrm{Re}E_\nu/\hbar+\Delta$, is linearly ramped within $0.5\,\mu$s from zero to maximum intensity $I_\mathrm{max}$, at which it is left for times on the order of $100\,\mu$s before being ramped back.
Corresponding to the experimental measurements we determined the fraction of atoms $[N_\mathrm{c}(t)-N_\mathrm{c}(t_0)]/N_\mathrm{c}(t_0)$ remaining in the condensate at time $t$ during the laser pulse.
The time evolution of this fraction is shown in Figure \ref{fig:condtimedep}.
We found that the final downward ramp of the laser power has no significant influence on the remaining condensate fraction.
Fig.~\ref{fig:condtimedep} shows the remaining fraction obtained from the non-Markovian non-linear Schr\"odinger equation (\ref{eq:NMNLSE}) (NMNLSE), for different maximum intensities $I_\mathrm{max}$, and the laser tuned to resonance, $\Delta=\Delta_0(I_\mathrm{max})=-164\,$MHz/(kW/cm$^2$)$\times I_\mathrm{max}$.
The intensity $I_\mathrm{max}=1.2\,$kW/cm$^2$ corresponds to the maximum intensity reported in Ref.~\cite{McKenzie02}.
At this intensity the condensate evolution can not be distinguished from the time dependence
\begin{align}
  \label{eq:solGPE}
  n_\mathrm{c}(\mathbf{x},t)
  &= \frac{n_\mathrm{c}(\mathbf{x},0)}
    {1+K_0(I_{\rm max},\Delta)n_\mathrm{c}(\mathbf{x},0)t},
\end{align}
corresponding to the solution of the GPE with a complex scattering length $a(I_\mathrm{max},\Delta_0(I_\mathrm{max}))$, Eq.~(\ref{eq:aofIandomega}), which gives a loss rate $K_0(I_{\rm max},\Delta_0)=(8\pi\hbar/m)|\Im a(I_{\rm max},\Delta_0(I_\mathrm{max}))|$.
Hence, the analysis in Ref.~\cite{McKenzie02} of their experimental data, in terms of the Gross-Pitaevskii theory, is consistent with our findings in the framework of the full many body dynamical theory.
This justifies, a posteriori, to fit our minimal model Hamiltonian to experimentally determined quantities as discussed in Appendix \ref{app:seppot}.

\subsubsection{Rate limits for condensate loss}
\label{sec:condratelimit}
Stimulated by recent theoretical work \cite{Kostrun00} experimental studies of photoassociation have focused on maximum rates at which, in principle, condensate atoms may be associated to molecules.
PA of Bose condensed atoms was, in the past, detected as loss from the atomic condensate fraction of the gas.

The upper limit to the possible association rate is set by the unitarity constraint to two body scattering \cite{Bohn99}, which may be expressed as the condition that the modulus squared of the inelastic scattering matrix element may not exceed one.
According to Ref.~\cite{McKenzie02}, the density and temperatures in the experiment were, however, such, that this limit had by far not been reached.

Based on a model of coupled mean fields for the atomic and molecular fractions a different limit to the condensate loss rate was proposed in Refs.~\cite{Kostrun00,Javanainen02}.
The theory presented in this work includes a process through which the molecules, which are identified with the closed channel bound states, dissociate into the continuum of background channel scattering states.
In a BEC this limit is of the order of $K_\mathrm{J}\sim(\hbar/m)n^{-1/3}$, i.e.~proportional to the mean distance between atoms.
For the experiment in Ref.~\cite{McKenzie02} $K_\mathrm{J}$ is far below the unitarity limit, and the intensity at which the loss rate is predicted to saturate due to this limit \cite{McKenzie02} coincides with the maximum intensity reported, of about $I=1.2\,$kW/cm$^2$.

We have investigated a range of intensities above $I=1.2\,$kW/cm$^2$, up to $I=50\,$kW/cm$^2$.
Fig.~\ref{fig:condtimedep}a shows the time evolution of the condensate fraction for on-resonance photoassociation pulses with $I_\mathrm{max}=10\,$kW/cm$^2$ and $I_\mathrm{max}=50\,$kW/cm$^2$, together with the corresponding time evolution according to the GPE.

We find that, according to our theory which fully takes into account the coupling into the continuum of excited states, the loss of condensate atoms does not saturate directly above $I_\mathrm{max}\sim1\,$kW/cm$^2$.
Nevertheless, the remaining fraction at the final time of the pulse is significantly reduced compared to the loss derived from the GPE.

Already at short times, the loss curves significantly deviate from the GPE result.
These deviations are due to the non-Markovian character of the evolution equation (\ref{eq:NMNLSE}).
The on-resonance scattering length (\ref{eq:aofIandomega}) is, for high intensities, dominated by its imaginary part $\mathrm{Im}\,a(I,\Delta_0(I))=-2a_\mathrm{bg}\delta(I)/\gamma$.
The Markov time of the many body evolution is of the order of the duration of a two body collision which is comparable to $m|a|^2/\hbar$.
For high light intensities, where $|a|$ becomes large, this timescale is significantly larger than the time over which the condensate mean field changes.

In Fig.~\ref{fig:locdecayrate} the non-Markovian character of the mean field evolution emerges as a temporally varying decay `rate'    
\begin{align}
  \label{eq:Kt}
  K(t)
  &= -\dot{N}_\mathrm{c}(t)\left[
     \int d\mathbf{x}\,n_\mathrm{c}^2(\mathbf{x},t)\right]^{-1}.
\end{align}
In the Gross-Pitaevskii theory the nonlinear interaction term is local in time such that $K$ is a constant, as indicated by the horizontal lines in Fig.~\ref{fig:locdecayrate}.
As soon as the interactions are resonantly enhanced, at around $t=0.5\,\mu$s, where the intensity dependent shift $\Delta_0(I)$ starts to compensate the laser detuning $\Delta$, the rate starts to rise.
Initially, however, $K(t)$ rises itself at a maximum rate (note the logarithmic time scale).
Whereas for $I_\mathrm{max}=1.2\,$kW/cm$^2$ the rate $K(t)$ quickly saturates at the Gross-Pitaevskii value $K_0$, for higher intensity the non-Markovian time evolution induces larger fluctuations before $K(t)$ settles in at a final value.
In view of these oscillations it is not justified to characterize the decrease by a single loss rate.
In Sect.~\ref{sec:collex} we shall discuss in more detail the physical grounds for these fluctuations.

Fig.~\ref{fig:locdecayrate} allows to relate our results to the limit proposed in Refs.~\cite{Kostrun00,Javanainen02}.
Inserting the local density $n_\mathrm{c}(\mathbf{R},t)$, calculated with Eq.~(\ref{eq:NMNLSE}), into $K_\mathrm{J}(\mathbf{R},t)\equiv(\hbar/m)[n_\mathrm{c}(\mathbf{R},t)]^{-1/3}$ we calculated the limit $K(t)$ given by Eq.~(\ref{eq:Kt}), with $\dot{N}_\mathrm{c}(t)=-\int\,d\mathbf{R}\,K_\mathrm{J}(\mathbf{R},t)n_\mathrm{c}(\mathbf{R},t)$, for a maximum intensity of $I_\mathrm{max}=50\,$kW/cm$^2$.
The values of $K(t)$ allowed according to this limit are indicated by the hatched area in Fig.~\ref{fig:locdecayrate}.
Even for such high intensities for which the condensate density decreases rapidly, thereby raising the upper limit $K_\mathrm{J}$, we find that the loss determined by Eq.~(\ref{eq:NMNLSE}) considerably exceeds this limit.

In summary: At the light intensities applied in the experiment in Ref.~\cite{McKenzie02} the loss of atoms from the condensate may be described, to a good approximation, by the GPE with a complex coupling constant proportional to the scattering length given in Eq.~(\ref{eq:aofIandomega}).
At higher intensities, the dynamics may no longer be described by a single rate constant.
Moreover, the fraction of atoms lost does not increase as strongly as predicted by the Gross-Pitaevskii theory.
However, the limitation of the loss rate is, according to our results, not as stringent as predicted in Refs.~\cite{Kostrun00,Javanainen02}.

\begin{figure}[tb]
\begin{center}
\includegraphics[width=0.45\textwidth]{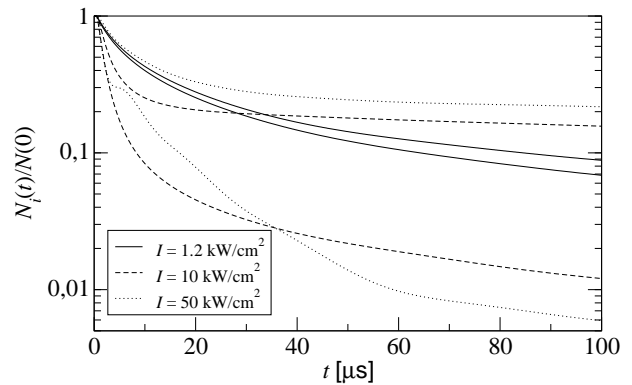}
\end{center}
\vspace*{-3ex}
\caption{
Time evolution of the fractions $N_i(t)/N(0)$ of atoms remaining in the condensate ($i=\,\,$`c') and in total ($i=\,\,$`tot'), of the initial atom number $N(0)\!\!=\,$4.0$\,\cdot\,$10$^{6}$ during a $100\,\mu$s pulse, with peak intensities of $1.2\,$kW/cm$^2$ (solid lines), $10\,$kW/cm$^2$ (dashed lines), and $50\,$kW/cm$^2$ (dotted lines). 
The fractions $N_i(t)/N(0)$ are given on a logarithmic scale.
The respective higher lying lines correspond to the total atom number, the lower ones to the condensate fraction.
}
\label{fig:CondvsNtottimedep}
\end{figure}
\subsubsection{Dissipative molecule formation}
\label{sec:molformation}
The results of the previous section have shown that in the vicinity of a photoassociative scattering resonance the condensate dynamics may deviate significantly from basic mean field behaviour, in particular at light intensities where the modulus of the scattering length (\ref{eq:aofIandomega}) is significantly enhanced.
In Section \ref{sec:resboundstates} we considered in detail the properties of the discrete near resonant dressed state with the highest energy $\mathrm{Re}E_\mathrm{d}$.
We shall now show how properties of this state mani\-fest themselves in the many body dynamics of near resonant photoassociation.
An observable of interest in this respect is the time evolution of the total number $N_\mathrm{tot}(t)$.
We recall that this number may decrease due to the finite life time $\gamma^{-1}$ of the closed channel state.

\begin{figure}[tb]
\begin{center}
\includegraphics[width=0.45\textwidth]{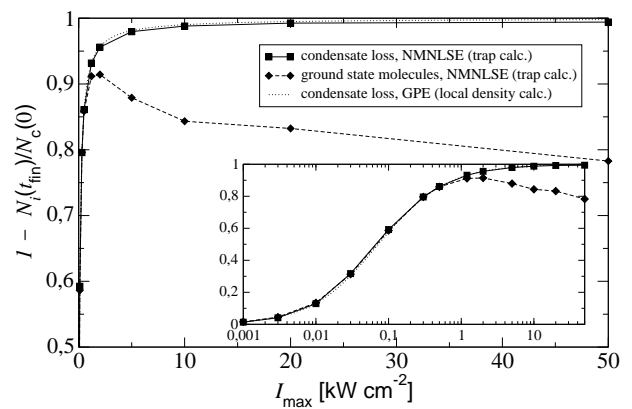}
\end{center}
\vspace*{-3ex}
\caption{
Loss fractions $1-N_{i}(t_{\rm fin})/N_{\rm c}(0)$, $i=$ `c', `tot', of the condensate (filled squares) and total atom (filled diamonds) loss at the end of the laser pulse, $t_{\rm fin}=\,$100$\,\mu$s, for different laser intensities $I_{\rm max}$.
The fraction indicated by the diamonds corresponds to the {\em number of ground state molecules} formed.
The solid line shows the result of a solution of the GPE in local density approximation.
The inset shows the same quantities on a logarithmic scale.
The modifications of the curves compared to Fig.~3 of Ref.~\cite{G0104} arise from taking into account, here, initial correlations as described in Sect.~\ref{sec:inicor}.
}
\label{fig:NlossofI}
\end{figure}
Using the solutions of Eq.~(\ref{eq:NMNLSE}) for different maximum laser intensities $I_\mathrm{max}$ and detunings $\Delta=\Delta_0(I_\mathrm{max})$, as well as Eq.~(\ref{eq:inicor}) and the two body time evolution $U_\mathrm{2B}$  determined with the minimal model Hamiltonian given in Appendix \ref{app:seppot} we calculated $N_\mathrm{tot}(t)$ from Eq.~(\ref{eq:dNndt}).
Fig.~\ref{fig:CondvsNtottimedep} shows the fraction $1-N_\mathrm{tot}(t)/N_\mathrm{tot}(0)$ of atoms remaining at time $t$ of the initial total number of  $N_\mathrm{tot}(0)=N_\mathrm{c}(0)=4.0\cdot10^6$ sodium atoms, for the same light intensities as in Fig.~\ref{fig:condtimedep}.
For comparison we add the condensate time evolution from Fig.~\ref{fig:condtimedep}, which corresponds to the respective lower lines.

Fig.~\ref{fig:NlossofI} shows the final fraction $1-N_\mathrm{tot}(t_\mathrm{fin})/N_\mathrm{tot}(0)$ of atoms remaining at $t=t_\mathrm{fin}=100\,\mu$s (filled diamonds, linearly interpolated by a dashed line).
Again, we compare this fraction to the final condensate fraction $1-N_\mathrm{c}(t_\mathrm{fin})/N_\mathrm{c}(0)$.
At low intensities the two fractions are identical, i.e.~all atoms missing from the condensate have been lost through the resonantly coupled unstable closed channel molecular state.
However, when increasing the intensity above ca.~$1\,$kW/cm$^2$ the total loss of atoms saturates and decreases slightly at the highest intensities.
According to Eqs.~(\ref{eq:Gammabgcl}) and (\ref{eq:ntot}) the loss of condensate atoms is due to the formation of binary correlations in the gas. 

It was shown in Refs.~\cite{KGB03,GKGJT04}, for the formally similar case of a Feshbach resonance, that the pair functions $\Phi_\mathrm{bg}$, $\Phi_\mathrm{cl}$ contain the information about the population of the background and closed channel components of the two body bound and scattering states.
With the corresponding quantum mechanical wave functions serving as form factors the pair functions yield observables for the number of molecules and excited atom pairs within the many body emsemble.
These observables were measured in a number of recent experiments where molecules and excited or `burst' atoms are produced using Feshbach resonances \cite{KGB03,KGJB03,KGG03,GKGJT04}.

In contrast to Feshbach resonances, a meaningful observable measuring the number of molecules produced in single color PA is neither given by the number of atom pairs in the dressed state with the highest energy $\mathrm{Re}E_\mathrm{d}$ which has been discussed in Sect.~\ref{sec:resboundstates}, nor by its closed channel component (cf.~Sect.~\ref{sec:bgChPop}).
Due to the fact that the decay time of the closed channel state is much shorter than any other timescale near resonance, any molecules formed are bound states of the background channel ground state potential.
Since we do not explicitly keep track of the target states of this decay process a meaningful observable for the maximum number of such molecules formed is the decrease of the total number of atoms $N_\mathrm{tot}$.
The diamonds in Fig.~\ref{fig:NlossofI} may therefore be seen as marking an upper limit to the number of molecules formed from the condensate atoms, which saturates at intensities above $I_\mathrm{max}\sim1\,$kW/cm$^2$.

We compare Fig.~\ref{fig:NlossofI} with the intensity dependence of the discrete dressed state close to threshold, shown in Figs.~\ref{fig:bgChPopofI} and \ref{fig:EdonResofI}.
As expected, the decreasing significance of the closed channel component of the dressed state above ca.~$I_\mathrm{max}=1\,$kW/cm$^2$ results in a reduction of the efficiency with which the condensate atoms may be associated to ground state molecules. 
The two body correlations induced by the high intensity laser partly relate to dressed two body states with a strongly enhanced lifetime.
According to Fig.~\ref{fig:EdonResofI} the lifetime of the dressed state at $I_\mathrm{max}=10\,$kW/cm$^2$ is on the order of $16\,\mu$s, a factor of 2000 larger than the lifetime of the bare closed channel vibrational state $|\phi_\nu\rangle$.

While the loss of atoms by spontaneous emission is reduced at high intensities, the laser stimulated loss to excited non-condensed atom pairs is gaining weight.
This explains why the saturation in the total atom loss is not accompanied by an equivalent saturation in the condensate loss.
Despite the fact that the closed channel component of the near threshold dressed state becomes negligible the resonant coupling to this state still allows for the population of the scattering states of atoms asymptotically in the internal ground state.
This transfer to the non-condensed fraction of the gas is a characteristic result of the full dynamical many body theory described in this article.

\begin{figure}[tb]
\begin{center}
\includegraphics[width=0.45\textwidth]{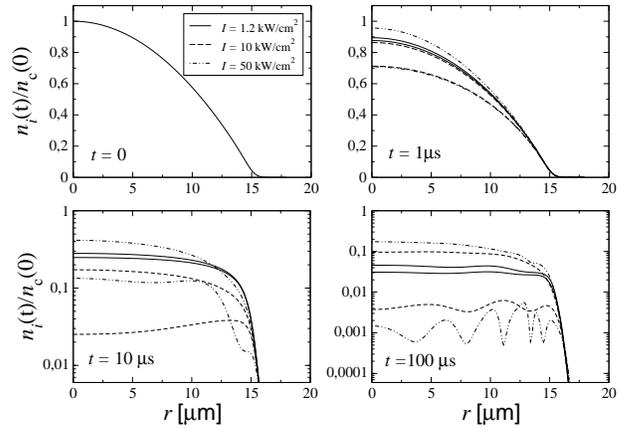}
\end{center}
\vspace*{-3ex}
\caption{
Time evolution of the radial density profile $n_i(r,t)/n_\mathrm{c}(0,0)$ of the fraction of condensate atoms ($i=$c) and the total number ($i=$tot) remaining at times $t=1,10,100\,\mu$s, for three different intensities $I=1.2$ (solid), $10$ (dashed), and $50\,$kW/cm$^2$ (dotted lines).
At time $t=0$ all evolutions start with a condensate of $4\cdot10^6$ sodium atoms in a spherically symmetric, harmonic $198\,$Hz trap. 
The respective upper curves correspond to the total atomic density distribution $n_\mathrm{tot}$, the lower curves to the condensate profile $n_\mathrm{c}$.
All profiles are normalized to the initial peak density $n_\mathrm{peak}=n_\mathrm{c}(0,0)$.
Note the logarithmic scale in the lower diagrams.
}
\label{fig:DensityProfile}
\end{figure}
\subsubsection{Collective excitations}
\label{sec:collex}
We would finally like to present results for the time evolution of the density profile of the atom cloud trapped in a spherically symmetric harmonic potential.
For the calculations presented in the previous sections we have taken into account, in Eq.~(\ref{eq:NMNLSE}), a trapping potential $V_\mathrm{trap}=\omega_\mathrm{ho}^2R^2$, with $\omega_\mathrm{ho}=(2\pi)198\,$Hz, corresponding to the mean trapping frequency in the experiment of McKenzie et al.~\cite{McKenzie02}.
As we have seen in Figs.~\ref{fig:condtimedep}a, \ref{fig:locdecayrate}b the condensate time evolution exhibits considerable oscillations at higher light intensities.
A detailed study of these oscillations is beyond the scope of this paper and will be presented elsewhere.
The oscillations are, however, expected to be understood analogously to a similar effect in a Feshbach resonance experiment of the kind presented in Ref.~\cite{Donley02}.
It was shown in Refs.~\cite{KGB03,KGJB03} that due to the fact that the condensate atoms and molecules are not completely orthogonal to each other close to a Feshbach resonance, the corresponding fractions of the gas exhibit residual oscillations.
In the dynamical evolution presented in Refs.~\cite{KGB03,KGJB03} these oscillations are seen during the evolution period between the initial and final Ramsey pulses.
Their frequency largely corresponds to the binding energy of the molecules formed on the positive scattering length side of the resonance.
Close to resonance, however, the experimental measurement of the oscillation frequency exhibited a shift \cite{Claussen03} which may be understood as caused by the many body environment in which the molecule association occurs \cite{Goralprivcomm}.
Our discussion of the near resonant dressed state spectrum in Sects.~\ref{sec:adiablevelsHom} and \ref{sec:adiablevelsTrap} has shown that for the laser detuning $\Delta=\Delta_0(I_\mathrm{max})$ during the pulse the real part of the uppermost dressed state is positive, of the order of a hundred kilohertz for $I_\mathrm{max}=10\,$kW/cm$^2$.
The time evolution of the rate shown in Fig.~\ref{fig:locdecayrate}b thus indicate a many body induced frequency shift which enables oscillations despite a positive dressed state energy. 

Fig.~\ref{fig:DensityProfile} shows the radial density profile of the condensate fraction, $|\Psi(R,t)|^2$, and of the total number of atoms, $n_\mathrm{tot}(R,t)$, for three different intensities $I_\mathrm{max}$ of the light, at various times during the light pulse.
The density of both, the condensate and the total atom number develops spatial oscillations, especially at the highest intensities.
These collective excitations arise because the frequency of the temporal oscillations depend on the local density of the cloud, which provides a further indication of their many body origin.
Our results suggest that an experiment with a Ramsey interferometer like setup as reported in Ref.~\cite{Donley02} may be used for precise molecular spectroscopy and to study the influence of the many body environment of the ultracold quantum degenerate gas.

\section{Conclusions}
\label{sec:concl}
A dynamical theory of photoassociation (PA) in a Bose-Einstein condensate has been presented which sets us in a position to study situations which due to the external conditions are no longer correctly described in the Gross-Pitaevskii approach.
The single color PA process type which is the subject of this article is described in a two-channel picture, accounting for an open background scattering channel and a molecular vibrational state in a closed channel potential of interacting electronically excited atoms.
The two-channel description was chosen since it allows to include the decay of the closed channel state due to spontaneous photon emission and its consequences for the many body evolution in a straightforward and convenient way.
We point out, though, that a single channel picture in principle suffices to obtain accurate results for the many body dynamics of the condensate fraction, as has been shown in Refs.~\cite{KGB03,KGG03} for the case of a Feshbach resonance.  

The modifications of the discrete spectrum of binary bound states near the photoassociation resonance have been described and the r\^{o}le of the finite liferime of the closed channel state has been shed light on.
We have extended our minimal Hamiltonian description of the low energy two-body scattering properties to the case where the atom pair is additionally confined by a tight spherically symmetric harmonic trap.
As an example, the adiabatic level structure near a Feshbach resonance as obtained from this approach was compared to corresponding results in Ref.~\cite{Bolda02} from numerical multichannel calculations and an effective scattering length model.
Excellent agreement confirmed the validity of the approach developed here.
By applying the tight trap Hamiltonian to single color PA we could illustrate the level structure behind the near resonance dressed state behavior found in the homogeneous case.
The real part of the dressed state energy was shown to intrude the positive real energy continuum of scattering states by crossing a number of low energy scattering levels before settling in and adiabatically going over into a scattering state itself.
Thereby the scattering level energies become complex close to the crossings, within a window of the order of the trap level spacing.
In other words, when tuning through the resonance the uppermost dressed level may survive on the side where the real part of its energy becomes positive, thereby avoiding a crossing of the scattering levels only in their energies' imaginary part.
We have discussed, furthermore, universal properties of the dressed state closest to the open channel threshold and shown that the spatial extent of its wave function on resonance scales, for high laser intensities $I$, quadratically with $I$.
This is a property specific to the case of an unstable closed channel state, i.e., it is not present in the bound state near a Feshbach resonance whose binding length scales with the scattering length and therefore linearly with the width of the resonance.

On the basis of the two body description developed here we have studied the dynamical evolution of a trapped condensate of sodium atoms whose interactions are strongly modified by a laser which is rapidly switched to a high light intensity, comparable to the situation in the experiment in Ref.~\cite{McKenzie02}.
We have computed the temporal evolution of the condensate fraction as well as of the total atom number for different intensities at the verge and above those reached in experiment.
The results exhibit the significance of the beyond mean field approach.
A simple loss rate description as in the Gross-Pitaevskii picture is not possible at high light intensities, and the amount of atoms lost within a certain time interval from the condensate is considerably limited compared to what may be expected from two-body theory.
Hence, our results corroborate earlier predictions of such a rate limit, but do not confirm the predicted value of the limit by allowing a faster loss.
We conclude that at the intensities reached in the experiment of McKenzie et al.~\cite{McKenzie02} a saturation of the condensate loss could not have been observed, but should be found if the intensity could be increased by roughly an order of magnitude.
Similar studies for other PA resonances are beyond the scale of this paper but may reveal saturation due to many body effects at more realistic light intensities.
The predicted condensate loss rate limitation is due to incoherent production of excited atom pairs, of the kind observed as a burst in experiments with Feshbach resonances, cf., e.g., Ref.~\cite{Donley02}.

We have also found a saturation of the number of ground state molecules which may be formed in the spontaneous electronic deexcitation process of the decaying closed channel bound states.
This saturation may be understood as resulting from the towards high intensities decreasing significance of the closed channel part in the two component dressed state vector vs.~its open channel part.
Similarly as the uppermost bound state on the positive scattering length side of a Feshbach resonance \cite{KGJB03} the extremely long range dressed state lives almost exclusively in the open channel, rendering the closed channel population irrelevant when describing the dynamics in a two channel picture.
For high light intensities, the achievable number of molecules is predicted to be significantly below the number of lost condensate atoms.

We have finally illustrated the collective excitations formed in the atomic cloud by the high intensity laser pulse.
These arise from a temporal oscillation of the condensate density during the laser illumination whose frequency depends on the local value of this density.
These oscillations, the detailed analysis of which is a subject of future work, are expected to be caused by a mean field shift of the dressed state energy and to be understood similarly as the residual coherent oscillations between condensate atoms and condensed molecules as observed in the Feshbach Ramsey interferometry \cite{Donley02} and described in Ref.~\cite{KGJB03}.

\acknowledgments
\noindent
I am very grateful to J\"urgen Berges, Joachim Brand, Keith Burnett, Olivier Dulieu, Krzysztof G\'oral, Paul Julienne, Thorsten K\"ohler, Markus Oberthaler, Robert Roth, Peter Schmelcher, Michael Schmidt, J\"org Schmiedmayer, Matthias Weidem\"uller, and Roland Wester for many stimulating discussions. 
This work has been supported by the Deutsche Forschungsgemeinschaft and the Alexander von Humboldt Foundation.

\begin{appendix}
\section{A minimal model for two body scattering}
\label{app:seppot}
\noindent
The scattering of a pair of atoms at very low kinetic energies, as in a Bose-Einstein condensed ensemble, is, to a very good accuracy, determined by a few parameters only.
As discussed in detail in Sects.~\ref{sec:2chdressed} a complete set of such parameters fixing a minimal two channel model Hamiltonian is given by the background scattering length $a_\mathrm{bg}$, the energy $E_{-1}$ of the last vibrational level in the background potential, the width $\delta(I)$ and shift $\Delta_0(I)$ of the laser induced resonance, and the decay width $\gamma$ of the resonantly coupled closed channel state.
Any Hamiltonian which reproduces these quantities may serve as a precise model to describe the full low energy scattering.

In this appendix we recall the basic ideas of the separable model introduced and discussed in detail in Refs.~\cite{KGB03,GKGJT04} and present the separable parameters used throughout this work to calculate both the static and dynamic scattering properties in Sects.~\ref{sec:2bodytheory} and \ref{sec:manybodytheory}.
An extension for the case of scattering in a spherical trapping potential follows which will be used in Appendix \ref{app:EigenstatesNHOP} to study the adiabatic energy level structure.
Finally the method for determining the two body time evolution will be described.

\subsection{Two channel model for asymptotically free atoms}
\label{app:seppotHom}
\noindent
Separable models are related to the observation that in certain scattering problems the transition matrix is to a very good approximation determined by a single pole \cite{Lovelace64,KGB03}.
For instance, in the case of low energy scattering considered here, this pole corresponds to the last vibrational state of the potential.
Knowledge of the potential and therefore of the spectrum of bound states would allow to calculate the $T$-matrix in this pole approximation.
While such a procedure would require an intermediate step with the much larger amount of information given by the potential the separable approach may be used in the opposite direction to fix an \emph{ansatz} for the last vibrational state of the potential $V$ in terms of a minimal set of defining parameters, cf.~Ref.~\cite{KGB03}.
Since the detailed form of the potential is irrelevant for collisions in ultracold gases, the low energy scattering properties are fully determined by this set.

In the single channel description of a scattering resonance presented in Ref.~\cite{KGB03} it is sufficient to provide the bound state energy and the scattering length for any configuration close to and away from the resonance.
The results of Ref.~\cite{KGB03} have demonstrated that such a single channel description is sufficient to reproduce the relevant low energy scattering properties and physical phenomena in the vicinity of a Feshbach resonance.
In Refs.~\cite{KGG03,GKGJT04} the single channel approach has been compared to a two channel description of a Feshbach resonance, and their results confirm that a single channel description suffices for the applications considered.

For practical convenience we will use a two channel description for single colour photoassociation.
The reasons for this are the finite life time of the closed channel state and the temporal variation of the light intensity and therefore of the width and shift of the resonance.
Thereby the two channel model automatically provides the correct dependence of the complex energy of the last vibrational state on detuning and intensity of the laser.
Our model consists of a separable ansatz for the background potential,
\begin{align}
  \label{eq:Vbgsep}
  V_\mathrm{bg}
  &=|\chi_\mathrm{bg}\rangle\xi_\mathrm{bg}\langle\chi_\mathrm{bg}|,
\end{align}
and the laser coupling
\begin{align}
  \label{eq:Wphinusep}
  W|\phi_\nu\rangle
  &=|\chi\rangle\zeta\sqrt{I},
\end{align}
with real parameters $\xi_\mathrm{bg}$ and $\zeta$.
Since the resonance shift $\Delta_0(I)$ and width $\delta(I)$ (cf.~Eq.(\ref{eq:aofIandomega})) depend linearly on $I$ at the intensities considered we have pulled a factor of $\sqrt{I}$ out of $\zeta$.

The practical convenience of the above ansatz results from choosing a Gaussian form for the states: 
\begin{align}
  \label{eq:chibg}
  \langle r|\chi_\mathrm{bg}\rangle
  &=\exp(-r^2/2\sigma_{\rm bg}^2)/(\sqrt{2\pi}\sigma_{\rm bg})^3
  \\
  \label{eq:chi}
  \langle r|\chi\rangle
  &=\exp(-r^2/2\sigma^2)/(\sqrt{2\pi}\sigma)^3.
\end{align}
Here, the parameters $\sigma_\mathrm{bg}$ and $\sigma$ define the width of the Gaussians and thus reflect the spatial extension of the van der Waals tail of the background potential and of the closed channel molecular state, respectively.

While the closed channel decay width $\gamma$ enters the Hamiltonian given in Eqs.~(\ref{eq:H2B0nu}) and (\ref{eq:V2Bnu}) as a real parameter, the four other parameters fix the separable ansatz of Eqs.~(\ref{eq:Vbgsep}) and (\ref{eq:Wphinusep}).
For the photoassociation resonance shown in Fig.~\ref{fig:multichannelsketch} the parameters assume the following values: $m\xi_{\rm bg}/(4\pi\hbar^2)=-191.4\,$a$_{\rm Bohr}$, $\sigma_{\rm bg}=23.97\,$a$_{\rm Bohr}$, $\xi=-(2\pi)57\,$GHz $(a_{\rm Bohr}^3/($kW$/$cm$^2))^{1/2}$, $\sigma=3.55\,a_{\rm Bohr}$, and $\gamma=(2\pi)18.5\,$MHz.
These values have been determined as follows.

Inserting the separable form (\ref{eq:Vbgsep}) into the Schr\"odinger equation for the last bound state of the background Hamiltonian yields an analytic expression of $E_{-1}$ in terms of $\xi_\mathrm{bg}$ and $\sigma_\mathrm{bg}$:
Multiplying the integral form $|\phi_{-1}\rangle=G_0(E_{-1})V_\mathrm{bg}|\phi_{-1}\rangle$ of the Schr\"odinger equation by $\langle\chi_\mathrm{bg}|$ from the left gives
\begin{align}
  \label{eq:EminusoneitoG0}
  1-\xi_\mathrm{bg}\langle\chi_\mathrm{bg}|G_0(E_{-1})|\chi_\mathrm{bg}\rangle
  &= 0,
\end{align}
The separable ansatz allows to calculate the expectation values of $G_0$ with respect to the states $|\chi_\mathrm{bg}\rangle,|\chi\rangle$ as
\begin{align}
  \label{eq:G0Eab}
  \langle\chi_a|G_0(E)|\chi_b\rangle
  &= \frac{m}{4\pi\hbar^2\sigma_{ab}}
  \left[x_{ab}e^{x_{ab}^2}\mathrm{erfc}(x_{ab})-\frac{1}{\sqrt{\pi}}\right].
\end{align}
Here $\chi_a,\chi_b\in\{\chi_\mathrm{bg},\chi\}$, $\sigma_{ab}=\sqrt{(\sigma_a^2+\sigma_b^2)/2}$, $\sigma_a,\sigma_b\in\{\sigma_\mathrm{bg},\sigma\}$, and $\mathrm{erfc}(x)=1-2\pi^{-1/2}$ $\int_0^xdu\,\exp[-u^2]$ is the complimentary error function of the argument $x_{ab}=\sqrt{-mE}\sigma_{ab}/\hbar$.

A second relation between the background scattering length, the energy $E_{-1}$, and the separable parameters $\xi_\mathrm{bg}$ and $\sigma_\mathrm{bg}$ is obtained from the background $T$-matrix which, in the separable potential approach takes the form \cite{KGB03}
\begin{align}
  \label{eq:TMatbgSepPot}
  T_\mathrm{bg}(z)
  &=\frac{|\chi_\mathrm{bg}\rangle\xi_\mathrm{bg}\langle\chi_\mathrm{bg}|}
    {1-\xi_\mathrm{bg}\langle\chi_\mathrm{bg}|G_0(z)|\chi_\mathrm{bg}\rangle}.
\end{align}
Using the definition $4\pi\hbar^2a_\mathrm{bg}/m=(2\pi\hbar)^3\langle0|T_\mathrm{bg}(0)|0\rangle$ of the background scattering length in terms of the background $T$-matrix $T_\mathrm{bg}(z)=V_\mathrm{bg}+V_\mathrm{bg}G_0(z)V_\mathrm{bg}$ one obtains an expression of $a_\mathrm{bg}$ in terms of $\xi_\mathrm{bg}$ and $\langle\chi_\mathrm{bg}|G_0(0)|\chi_\mathrm{bg}\rangle$.
With Eq.~(\ref{eq:G0Eab}) this gives:
\begin{align}
  \label{eq:xibgitoabg}
  \xi_\mathrm{bg}
  &= \frac{4\pi\hbar^2a_\mathrm{bg}}{m}
     \left[1-\frac{a_\mathrm{bg}}{\sqrt{\pi}\sigma_\mathrm{bg}}\right]^{-1}.
\end{align}
This equation yields, inserted in Eq.~(\ref{eq:EminusoneitoG0}) and using Eq.~(\ref{eq:G0Eab}), a relation which determines $\sigma_\mathrm{bg}$ in terms of $a_\mathrm{bg}$ and $E_{-1}$:
\begin{align}
  \label{eq:sigmabgitoabgEbbg}
  1-\frac{a_\mathrm{bg}}{\sigma_\mathrm{bg}}xe^{x^2}\mathrm{erfc}(x)
  &= 0.
\end{align}
Having determined $\sigma_\mathrm{bg}$ in this way, Eq.~(\ref{eq:xibgitoabg}) finally yields $\xi_\mathrm{bg}$.

Introducing the pole approximation with respect to the closed channel state $|\phi_\nu\rangle$ we have already obtained the off diagonal coupling in a separable form.
The parameters $\xi$ and $\sigma$ are therefore directly determined from Eqs.~(\ref{eq:deltaitoME}) and (\ref{eq:shiftitoME}). 
Using the Lippmann-Schwinger equation
\begin{align}
  \label{eq:LippmannSchwinger}
   |\phi^{(+)}_\mathbf{p}\rangle
   &= |\mathbf{p}\rangle+G_0(p^2/m+i0)T_\mathrm{bg}(p^2/m+i0)|\mathbf{p}\rangle
\end{align} 
for the scattering states in the background channel and the expression 
\begin{align}
  \label{eq:GbgitoG0Tbg}
  G_\mathrm{bg}(z)
  &= G_0(z)[1+T_\mathrm{bg}(z)G_0(z)]
\end{align}
for the background channel Green's function, Eq.~(\ref{eq:Gbgofz}), in terms of the background channel $T$-matrix, Eq.~(\ref{eq:TMatbgSepPot}), one obtains:
\begin{align}
  \label{eq:sigmaitoDeltaandGamma}
  \frac{\partial\Delta_0/\partial I}{\partial\delta/\partial I}
  &= \frac{1}{\sqrt{\pi}\sigma}\frac{1-
     \frac{a_\mathrm{bg}}{\sqrt{\pi}\sigma}
     \frac{2}{1+(\sigma_\mathrm{bg}/\sigma)^2}}
     {\left[1-
     \frac{a_\mathrm{bg}}{\sqrt{\pi}\sigma}
     \sqrt{\frac{2}{1+(\sigma_\mathrm{bg}/\sigma)^2}}\right]^2}
  \\
  \label{eq:xiitoDeltaandGamma}
  \zeta
  &= \frac{\sqrt{\frac{2\pi\hbar^3}{m}\frac{\partial\delta}{\partial I}}}
     {1-
     \frac{a_\mathrm{bg}}{\sqrt{\pi}\sigma}
     \sqrt{\frac{2}{1+(\sigma_\mathrm{bg}/\sigma)^2}}}.
\end{align}
To arrive at these relations we have also used Eq.~(\ref{eq:G0Eab}) and the approximative linearity in $I$ of $\delta$ and $\Delta_0$.
Eqs.~(\ref{eq:sigmaitoDeltaandGamma}) and (\ref{eq:xiitoDeltaandGamma}), together with Eqs.~(\ref{eq:xibgitoabg}) and (\ref{eq:sigmabgitoabgEbbg}) determine the parameters of the minimal model Hamiltonian given above in terms of experimentally known or otherwise theoretically determined quantities.

In Sect.~\ref{sec:resboundstates} we used the separable model to calculate the near resonant bound states and their energies.
As an example we give the equation which was used to determine the energy dependence of the dressed states on the intensity $I$ and detuning $\Delta$.
Using Eqs.~(\ref{eq:EnergyDressed}), (\ref{eq:Wphinusep}), (\ref{eq:TMatbgSepPot}), and (\ref{eq:GbgitoG0Tbg}) we find:
\begin{align}
  \label{eq:EditoSepPot}
  E_\mathrm{d}
  &=\Delta-i\gamma/2
   + \zeta^2I\Bigg[
    \langle\chi|G_0(E_\mathrm{d})|\chi\rangle
  \nonumber\\
  &\qquad+\frac{
    \xi_\mathrm{bg}(\langle\chi|G_0(E_\mathrm{d})|\chi_\mathrm{bg}\rangle)^2}{
   1-\xi_\mathrm{bg}
    \langle\chi_\mathrm{bg}|G_0(E_\mathrm{d})|\chi_\mathrm{bg}\rangle}\Bigg].
\end{align}
Here we have also used, that the \emph{time independent} left eigenstate $\langle\tilde\phi_\nu|$ may not be distinguished from the corresponding right eigenstate $\langle\phi_\nu|$, such that $\langle\tilde\phi_\nu|W=\zeta\sqrt{I}\langle\chi|$ (cf.~Appendix \ref{app:EigenstatesNHOP}).
The expectation values of the Green's function $G_0$ may be determined using Eq.~(\ref{eq:G0Eab}).

\subsection{Extension to trapped atoms}
\label{app:seppotTrap}
\noindent
Within an external trapping potential the relative motion of two atoms is additionally restricted with the effect that the spectrum of positive energy scattering state is no longer continous.
In this section we introduce a separable potential approach to treat low energy two body scattering within a confining spherically symmetric harmonic trap.
For the purpose of this article our goal is to calculate the spectrum of negative and positive energy dressed states of the relative motion within the trap and their dependence on the detuning and intensity of the laser.
To this end we will fix the separable parameters such that in the homogeneous limit they yield the known low energy scattering properties.

\subsubsection{Two channel separable potential}
\label{app:twochseppotTrap}
\noindent
An effective scattering length model of ultracold atomic collisions in tight harmonic traps has been discussed in detail in Ref.~\cite{Bolda02}.
Like this model our minimal Hamiltonian model with separable interactions accurately reproduces the positive energy spectrum of the atomic relative motion determined from multichannel closed coupling calculations presented in \cite{Bolda02}.
We expect the accuracy of our model to hold for tight traps as long as the oscillator length considerably exceeds the length scale $l_\mathrm{vdW}$ of the van der Waals interactions. 

We consider two atoms whose relative motion in the background channel is, without the coupling laser, described by the Hamiltonian
\begin{align}
 \label{eq:H2B0nutrap}
  H_\mathrm{2B}
  &= H_0^\mathrm{ho}
   |\mathrm{bg}\rangle\langle\mathrm{bg}|
  +|\phi_\nu,\mathrm{cl}\rangle E_\nu \langle\tilde\phi_\nu,\mathrm{cl}|
  +V_\mathrm{2B}.
\end{align}
Here, the second term is equivalent to the corresponding term in Eq.~(\ref{eq:H2B0nu}), and $V_\mathrm{2B}$ is the two channel interaction picture potential matrix given in Eq.~(\ref{eq:V2Bnu}).
The Hamiltonian of trapped, non-interacting atom pairs reads, in the spatial representation:
\begin{align}
 \label{eq:Hbg0trap}
  H_0^\mathrm{ho}(r)
  &= -\frac{\hbar^2\nabla^2}{m} + \frac{1}{4}m\omega_\mathrm{ho}^2r^2,
\end{align}
where $\omega_\mathrm{ho}/2\pi$ is the harmonic oscillator frequency.
The motion of non-interacting atoms is restricted to the spherical trap modes $|n,l\rangle$, with
\begin{align}
  \label{eq:trapmodes}
  H_0^\mathrm{ho}|n,l\rangle
  &= E_{n,l}^{(0)}|n,l\rangle,
  \\
  \label{eq:trapenergies}
  E_{n,l}
  &= \hbar\omega_\mathrm{ho}\left[\frac{3}{2}+2n+l\right].
\end{align}
Since the trap is isotropic we may, as in the uniform case, consider only the $s$-wave term ($l=0$) in the expansion of the scattering amplitude.
We define an $s$-wave scattering length by 
\begin{align}
  \label{eq:abgtrap}
  a_\mathrm{bg}^\mathrm{ho}(E)
  &= \frac{m}{4\pi\hbar^2}{\cal N}^{-1}\langle 0,0|T_\mathrm{bg}(E)|0,0\rangle,
\end{align}
where $|0,0\rangle$ is the ($n=0,l=0$) oscillator ground state defined by Eq.~(\ref{eq:trapmodes}), and the $T$-matrix is evaluated at zero or positive energy $E\ge0$.
Furthermore, ${\cal N}=(2\pi)^{-3/2}l_\mathrm{ho}^{-3}$, determined by the oscillator length $l_\mathrm{ho}=\sqrt{\hbar/m\omega_\mathrm{ho}}$, is a normalization factor which is chosen such that in the homogeneous limit the scattering length $a_\mathrm{bg}^\mathrm{ho}(0)$ defined by Eq.~(\ref{eq:abgtrap}) is equivalent to the usual $a_\mathrm{bg}$.

We introduce a separable potential as in Eqs.~(\ref{eq:Vbgsep}), (\ref{eq:TMatbgSepPot}) and make the following convenient ansatz for the trap representation of the state $|\chi_\mathrm{bg}\rangle$ in terms of a real parameter $\sigma_\mathrm{bg}$:
\begin{align}
  \label{eq:chibgtrap}
  \langle n,0|\chi_\mathrm{bg}\rangle
  &= \left[\left(1+2n\right){\cal N}
     \left({n-\frac{1}{2}}\atop{n}\right)
     \right]^{1/2}e^{-\sigma_\mathrm{bg}^2/l_\mathrm{ho}^2}.
\end{align}
$({a\atop b})=\Gamma(a+1)/[\Gamma(b+1)\Gamma(a-b)]$ denotes the generalized binomial coefficient.

Using the spectral expansion
\begin{align}
  \label{eq:G0SpecExp}
  G_0(z)
  &= \sum_{n,l}|n,l\rangle(z-E_{n,l}^{(0)})^{-1}\langle n,l| 
\end{align}
of the free energy dependent Green's function its expectation values with respect to the states $|\chi_\mathrm{bg}\rangle,|\chi\rangle$ are (cf.~Eq.~(\ref{eq:G0Eab})): 
\begin{align}
  \label{eq:G0Eabtrap}
  &\langle\chi_a|G_0(E)|\chi_b\rangle
  \nonumber\\
  &\quad= -\frac{m}{4\pi\hbar^2\sqrt{\pi}l_\mathrm{ho}}
  \left[\sqrt{\frac{2}{1-\alpha_{ab}^2}}
  -\left(y-\frac{1}{2}\right)J(y,\alpha_{ab})\right].
\end{align}
Here, $y=-E/\hbar\omega_\mathrm{ho}$, $\alpha_{ab}=\exp[-\sigma_{ab}^2/l_\mathrm{ho}^2]$, $\sigma_{ab}=\sqrt{(\sigma_a^2+\sigma_b^2)/2}$, $\sigma_a,\sigma_b\in\{\sigma_\mathrm{bg},\sigma\}$, and the integral $J(y,\alpha)$ may be expressed in terms of a hypergeometric function:
%
\begin{align}
  \label{eq:Jyalpha}
  J(y,\alpha)
  &= \sqrt{2}\int_0^1du\frac{u^{1/2-y}}{\sqrt{1-u^2\alpha^2}}
    \nonumber\\
  &=\frac{2\sqrt{2}}{3-2y}
    F\left(\mbox{$\frac{1}{2},\frac{1}{2}(\frac{3}{2}-y);$}
           \mbox{$\frac{1}{2}(\frac{7}{2}-y);\alpha^2$}
    \right).
\end{align}

Inserting the separable $T$-matrix, Eq.~(\ref{eq:TMatbgSepPot}), into Eq.~(\ref{eq:abgtrap}) and using the ansatz (\ref{eq:chibgtrap}) we obtain (cf.~Eq.~(\ref{eq:xibgitoabg})): 
\begin{align}
  \label{eq:xibgitoabgtrap}
    \xi_\mathrm{bg}(E)
  &= \frac{4\pi\hbar^2a_\mathrm{bg}^\mathrm{ho}(E)}{m}
     \Bigg[1-\frac{a_\mathrm{bg}^\mathrm{ho}(E)}
                  {\sqrt{\pi}l_\mathrm{ho}}
           \sqrt{\frac{2}{1-e^{-2\sigma_\mathrm{bg}^2/l_\mathrm{ho}^2}}}
  \nonumber\\
  &\phantom{= \frac{4\pi\hbar^2a_\mathrm{bg}^\mathrm{ho}(E)}{m}}
   -\frac{1-2y}{2\sqrt{2}}J(y,e^{-2\sigma_\mathrm{bg}^2/l_\mathrm{ho}^2})
     \Bigg]^{-1}.
\end{align}
One may show that for any fixed $-y=E/\hbar\omega_\mathrm{ho}\ge0$, Eq.~(\ref{eq:xibgitoabgtrap}) is, in the weak trap limit $\omega_\mathrm{ho}\to0$, equivalent to Eq.~(\ref{eq:xibgitoabg}).
For the following considerations we define $\xi_\mathrm{bg}$ through Eq.~(\ref{eq:abgtrap}) with $E=\hbar\omega_\mathrm{ho}/2$.

Using Eq.~(\ref{eq:G0Eabtrap}) we then obtain a relation determining the spectrum of energies $E$ in the background channel analogous to Eq.~(\ref{eq:sigmabgitoabgEbbg}):
\begin{align}
  \label{eq:sigmabgitoabgEbbgtrap}
  1-\frac{a_\mathrm{bg}^\mathrm{ho}}{\sqrt{\pi}l_\mathrm{ho}}
    \left(\frac{1}{2}-y\right)\,J(y,e^{-\sigma_\mathrm{bg}^2/l_\mathrm{ho}^2})
  &= 0,
\end{align}
where $y=-E/\hbar\omega_\mathrm{ho}$, $a_\mathrm{bg}^\mathrm{ho}\equiv a_\mathrm{bg}^\mathrm{ho}(\hbar\omega_\mathrm{ho}/2)$.
Again, in the weak trap limit, $\omega_\mathrm{ho}\to0$, this relation is equivalent to Eq.~(\ref{eq:sigmabgitoabgEbbg}).
We therefore used Eqs.~(\ref{eq:xibgitoabg}) and (\ref{eq:sigmabgitoabgEbbg}) to fix the separable parameters $\xi_\mathrm{bg}$ and $\sigma_\mathrm{bg}$.

The separable parameters $\sigma$ and $\zeta$ (cf.~Eq.~(\ref{eq:Wphinusep})) are determined in the same way as in the uniform case.
Using the arguments which have lead to Eqs.~(\ref{eq:sigmaitoDeltaandGamma}), (\ref{eq:xiitoDeltaandGamma}) we find:
\begin{align}
  \label{eq:sigmaitoDeltaandGammatrap}
  \frac{\partial\Delta_0/\partial I}{\partial\delta/\partial I}
  &= \frac{|1+\xi_\mathrm{bg}
            \langle\chi|G_\mathrm{bg}(0)|\chi\rangle|^2}
	  {\langle\chi|G_\mathrm{bg}(0)|\chi\rangle},
  \\
  \label{eq:xiitoDeltaandGammatrap}
  \zeta
  &= \left[\frac{\partial\delta/\partial I}
     {\langle\chi|G_\mathrm{bg}(0)|\chi\rangle}\right]^{1/2}.
\end{align}
The expectation values of the Greens function $G_\mathrm{bg}(0)$ are evaluated using Eqs.~(\ref{eq:TMatbgSepPot}), (\ref{eq:GbgitoG0Tbg}), and (\ref{eq:G0Eabtrap}).

\subsubsection{Accuracy of separable approximation in tight traps}
\label{app:accuracyseppotTrap}
\begin{figure}[tb]
\begin{center}
\includegraphics[width=0.45\textwidth]{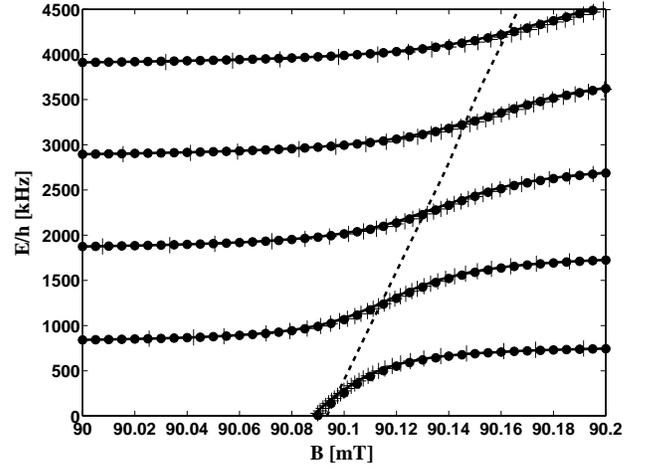}
\end{center}

\caption{
Comparison of adiabatic energy levels calculated in this paper with the result from Ref.~\protect\cite{Bolda02} for the five lowest positive energy levels near the $B_0=90.1\,$mT Feshbach resonance of $^{23}$Na atoms colliding in the $(F=1,m_F=+1)$ hyperfine ground state, which are confined in a $\nu_\mathrm{ho}=500\,$kHz spherically harmonic trap.
I thank P. Julienne for providing me with Fig.~4 of Ref.~\protect\cite{Bolda02}.
}
\label{fig:FBPosSpectrumTrap}
\end{figure}

\noindent
To check the accuracy of the minimal model introduced here we have considered the scattering of $^{23}$Na atoms in the $(F=1,m_F=+1)$ stretched state, in the vicinity of the Feshbach resonance at $B_0=90.1\,$mT for the case that the atoms are confined within a tight spherically symmetric trap with $\omega_\mathrm{ho}=(2\pi)500\,$kHz.
In Ref.~\cite{Bolda02} the five lowest positive energy levels were presented as a function of the magnetic field detuning from the Feshbach resonance, determined with multichannel close coupling methods from high precision data for the relevant interaction potentials.
Here, we use Eq.~(\ref{eq:EnergyDressed}), with $\Delta(B)=(dE_\nu/dB)(B-B_0)=(52.4\,$MHz$\,h/$mT$)(B-B_0)$ and $\gamma=0$, together with the separable ansatz (\ref{eq:chibgtrap}) for the interactions modified by the trap, to determine the spectrum of positive energy states.
Inserting the parameters $m\xi_{\rm bg}/(4\pi\hbar^2)=-184\,a_\mathrm{Bohr}$, $\sigma_\mathrm{bg}=26.5\,a_\mathrm{Bohr}$, $\zeta=-44.0\,$GHz$\,h\,a_\mathrm{Bohr}^{3/2}$ and $\sigma=17.4\,a_\mathrm{Bohr}$ which were determined from the low energy scattering properties of untrapped atoms close to the resonance (cf.~Ref.~\cite{KGG03}), we obtain the dependence of the energy levels on $B$. 
The results are shown in Fig.~\ref{fig:FBPosSpectrumTrap} (`+'-signs) where they are compared to the results from Ref.~\cite{Bolda02}, of numerical calculations (filled circles) as well as of an effective scattering length model (solid line).

In Ref.~\cite{Busch98} the energies $E$ of the relative motion of trapped atom pairs were found, using the pseudopotential approximation to the binary interactions, to be given by the solutions of the equation \cite{fn:reducedmassinholength}
\begin{align}
  \label{eq:WilkensInterceptEq}
  \frac{a}{l_\mathrm{ho}}
  &= \frac{1}{\sqrt{2}}\tan\left[\frac{\pi}{2}\left(\frac{1}{2}-y\right)\right]
     \frac{\Gamma\left(\frac{1}{4}-\frac{y}{2}\right)}
          {\Gamma\left(\frac{3}{4}-\frac{y}{2}\right)},
\end{align}
where $a$ is the scattering length entering the pseudopotential and $y=-E/\hbar\omega_\mathrm{ho}$.
Extending this approach, the authors of Ref.~\cite{Bolda02} argue that $a$ needs to be replaced by an effective scattering length $a_\mathrm{eff}(E)=-\tan[\delta_0(k)]/k$, where $E=\hbar^2k^2/m$ and $\delta_0(k)$ is the momentum dependent $s$-wave phase shift.
This causes the magnetic field value, where a particular energy level corresponds to a diverging scattering length, to be shifted for subsequent levels, as indicated by the slanted dashed line in Fig.~\ref{fig:FBPosSpectrumTrap}.
For the level structure determined from Eq.~(\ref{eq:WilkensInterceptEq}) the dashed line would be essentially vertical.

As discussed in detail in Refs.~\cite{KGB03,GKGJT04} the parameter $\sigma_\mathrm{bg}$ introduces the scale of the van der Waals length $l_\mathrm{vdW}$ into the description of the atomic interactions.
Hence the pseudopotential approximation should be recovered in the limit $\sigma_\mathrm{bg}\to0$.
Using the relation $F(a,b;c;1)=\Gamma(c)\Gamma(c-a-b)/[\Gamma(c-a)\Gamma(c-b)]$ between the hypergeometric function and Euler's $\Gamma$ function one may indeed show that in this limit Eq.~(\ref{eq:WilkensInterceptEq}) is equivalent to Eq.~(\ref{eq:sigmabgitoabgEbbgtrap}), identifying thereby $a=a_\mathrm{bg}^\mathrm{ho}$.

Hence, the excellent agreement of our results with the results of Ref.~\cite{Bolda02}, as seen in Fig.~\ref{fig:FBPosSpectrumTrap}, not only shows that the minimal model Hamiltonian remains applicable outside the Wigner law regime, where the scattering length is much larger than the harmonic oscillator length.
It also nicely illustrates that the particular trap induced level shifts close to resonance are strongly influenced by the far ranged van der Waals interactions.

\subsection{Determination of the time dependent coupling function}
\label{app:seppotTimeEvol}
\noindent
The essential ingredient of the non-Markovian non-linear equation (\ref{eq:NMNLSE}) for the mean field $\Psi(\mathbf{x},t)$ is the antiderivative $h(t,\tau)=(2\pi\hbar)^3\langle0,\mathrm{bg}|V_\mathrm{2B}(t)
U_\mathrm{2B}(t,\tau)|0,\mathrm{bg}\rangle\,\theta(t-\tau)$ of the expectation value of the time dependent $T$-matrix between the zero momentum plane wave states in the background channel.
\begin{align}
  \label{eq:TdepTM}
  &\partial h(t,\tau)/\partial t
  = g(t,\tau)
  \nonumber\\
  &\ = (2\pi\hbar)^3\langle0,\mathrm{bg}|
     V(t)[\delta(t-\tau)+G_\mathrm{2B}(t,\tau)V(\tau)]|0,\mathrm{bg}\rangle.
\end{align}
In the following we will briefly summarize the procedure involving the minimal model Hamiltonian introduced in Subsection \ref{app:seppotHom} which allows the time dependent coupling function $h(t,\tau)$ to be determined numerically.

The interaction picture time evolution operator
\begin{align}
  \label{eq:U2B}
  U_\mathrm{2B}(t,\tau)
  &={\cal T}e^{-\frac{i}{\hbar}\int_\tau^tdt'\,H_\mathrm{2B}(t')}
  \nonumber\\
  &\equiv (|\mathrm{bg}\rangle,|\mathrm{cl}\rangle)
      \left(\begin{array}{cc}
      U_{11}(t,\tau) & 
      U_{12}(t,\tau) \\
      U_{21}(t,\tau) & 
      U_{22}(t,\tau)
      \end{array}\right)
      \left(\!\!\begin{array}{c}
      \langle\mathrm{bg}|\\ 
      \langle\mathrm{cl}|
      \end{array}\!\!\right)
\end{align}
corresponding to the two channel Hamiltonian $H_\mathrm{2B}(t)$ in Eq.~(\ref{eq:TwoChSSE}) may be written in terms of the background and closed channel Green's functions of the uncoupled motion ($W=0$) by considering the off diagonal coupling as an inhomgeneous term in the Schr\"odinger equation.
The matrix elements relevant for our purposes are then determined by the following closed set of coupled integral equations: 
\begin{align}
  \label{eq:U2B11}
  U_{11}(t,\tau)
  &= U_\mathrm{bg}(t-\tau)
  \nonumber\\ 
  &\quad+ \int_\tau^tdt'\,G_\mathrm{bg}(t-t')
     W(t')P_\nu e^{i\varphi(t')}U_{21}(t',\tau),
  \\
  \label{eq:U2B21}
  U_{21}(t,\tau)
  &= \frac{1}{i\hbar}\int_\tau^tdt'\,e^{-i\varphi(t')}
     P_\nu W(t') U_{11}(t',\tau),
\end{align}
Here, $U_\mathrm{bg}(t)=\exp[-iH_\mathrm{bg}t/\hbar]$ is the background channel time evolution operator, $G_\mathrm{bg}(t)=-i\theta(t)U_\mathrm{bg}(t)/\hbar$ the corresponding Green's function of the Schr\"odinger operator $i\hbar\partial/\partial t-H_\mathrm{bg}$, and $\varphi(t)=-\int_{t_0}^td\tau(E_\nu-\hbar\omega)/\hbar$ is the phase angle describing the time evolution of the closed channel state $|\phi_\nu\rangle$ (cf.~Eq.~(\ref{eq:TwoChSSE})).
We have furthermore assumed that the detuning $\Delta=\omega-\mathrm{Re}E_\nu/\hbar$ of the laser remains constant in time.
While this assumption is chosen in accordance with the experimental sitation of Ref.~\cite{McKenzie02} and simplifies the expressions, the general case of varying detuning may be implemented straightforwardly.
 
In terms of the matrix elements $U_{11}$ and $U_{21}$ the coupling function $h(t,\tau)$ reads
\begin{align}
  \label{eq:hitoU2Bij}
  h(t,\tau)
  &= (2\pi\hbar)^3\theta(t-\tau)\Big[
     \langle0|V_\mathrm{bg}U_\mathrm{bg}(t-\tau)|0\rangle
  \nonumber\\
  &
  +\ \int_\tau^tdt'\frac{\partial}{\partial t}
   \langle0|U_\mathrm{bg}(t-t')W(t')P_\nu e^{i\varphi(t')}U_{21}(t',\tau)|0\rangle
   \Big].
\end{align}
To evaluate the integral equations (\ref{eq:U2B11})--(\ref{eq:hitoU2Bij}) we have used the separable ansatz for $V_\mathrm{bg}$ and $W|\phi_\nu\rangle$, Eqs.~(\ref{eq:Vbgsep}) and (\ref{eq:Wphinusep}), respectively.
We are thus led to the coupled set of integral equations
\begin{widetext}
\begin{align}
  \label{eq:IntEqforu11and21}
  \left(\begin{array}{c}
      u_{11}(t,\tau) \\
      u_{21}(t,\tau) 
  \end{array}\right)
  &=  
  \left(\begin{array}{c}
      u_\mathrm{bg}(t-\tau) \\
      0 
  \end{array}\right)
  -\int_\tau^tdt'\sqrt{I(t')}\left[\frac{\partial}{\partial t'}
  \left(\begin{array}{cc}
      0 & \tilde{h}_\mathrm{bg}(t-t') \\
      \tilde{h}_\mathrm{cl}(t-t') & 0
  \end{array}\right)\right]
  \left(\begin{array}{c}
      u_{11}(t',\tau) \\
      u_{21}(t',\tau) 
  \end{array}\right)
\end{align}
\end{widetext}
for the functions
\begin{align}
  \label{eq:u11}
  u_{11}(t,\tau)
  &= \zeta\langle\chi|U_{11}(t,\tau)|0\rangle,
  \\
  \label{eq:u21}
  u_{21}(t,\tau)
  &= e^{i\varphi(t)}\langle\phi_\nu|U_{21}(t,\tau)|0\rangle,
\end{align}
in terms of
\begin{align}
  \label{eq:ubg}
  u_\mathrm{bg}(t)
  &= \zeta\langle\chi|U_\mathrm{bg}(t)|0\rangle,
\end{align}
\begin{align}
  \label{eq:htildebg}
  (\partial/\partial t)\tilde{h}_\mathrm{bg}(t)
  &= \zeta^2\langle\chi|G_\mathrm{bg}(t)|\chi\rangle,
  \\
  \label{eq:htildecl}
  \tilde{h}_\mathrm{cl}(t)
  &= \frac{e^{i(\Delta+i\gamma/2)t}}{-\hbar(\Delta+i\gamma/2)}.
\end{align}
Inserting the separable ansatz into Eq.~(\ref{eq:hitoU2Bij}) the equation for the coupling function $h(t,\tau)$ reads
\begin{widetext}
\begin{align}
  \label{eq:h11}
  h(t,\tau)
  &= h_\mathrm{bg}(t-\tau)
   -\theta(t-\tau)\int_\tau^tdt'\sqrt{I(t')}u_\mathrm{21}(t',\tau)
    \frac{\partial}{\partial t'}\Big[\theta(t-t')u_\mathrm{bg}(t-t')\Big],
\end{align}
\end{widetext}
where
\begin{align}
  \label{eq:hbg}
  h_\mathrm{bg}(t)
  &= (2\pi\hbar)^3\theta(t)\langle0|V_\mathrm{bg}U_\mathrm{bg}(t)|0\rangle.
\end{align}
The calculation of the coupling function $h(t,\tau)$ therefore reduces to evaluating the functions $h_\mathrm{bg}$, $u_\mathrm{bg}$, and $\tilde{h}_\mathrm{bg}$.
Note that all what is needed of the information included in the full scattering potential $V_\mathrm{bg}$ are matrix elements between the Gaussian wave functions constituing the separable ansatz of Eqs.~(\ref{eq:Vbgsep}) and (\ref{eq:Wphinusep}).
Hence, the minimal set of parameters introduced in Subsection \ref{app:seppotHom} determines the full dynamics of the low energy scattering of a pair of atoms near the photoassociation resonance.

In all many body applications considered in this paper the background channel Hamiltonian does not change in time. 
Hence, the functions $h_\mathrm{bg}$, $u_\mathrm{bg}$, and $\tilde{h}_\mathrm{bg}$ only depend on a single time variable $t-\tau$ and may therefore be evaluated using the spectral decomposition of the time dependent Green's function,
\begin{align}
  \label{eq:Gbgspec}
  G_\mathrm{bg}(t)
  &= \frac{1}{i\hbar}\theta(t)\int d\mathbf{p}\,e^{-i(p^2/m)t/\hbar}
     |\phi_\mathbf{p}^\mathrm{bg}\rangle
            \langle\phi_\mathbf{p}^\mathrm{bg}|
  \nonumber\\
  &\quad+\sum_\lambda e^{i|E_\lambda|t/\hbar}
     \frac{|\phi_\lambda^\mathrm{bg}\rangle\langle\phi_\lambda^\mathrm{bg}|}.
\end{align}
Here, $|\phi_\mathbf{p}^\mathbf{bg}\rangle$ and $|\phi_\lambda^\mathbf{bg}\rangle$ are the background channel scattering and bound states, with eigenenergies $p^2/m$ and $E_\lambda$, respectively.
Taking only into account the contribution from the scattering states, i.e.~the first term on the right side of Eq.~(\ref{eq:Gbgspec}), we obtain
\begin{align}
  \label{eq:hbgspec}
  &h_\mathrm{bg}(t)
  = \theta(t)\Bigg[\frac{4\pi\hbar^2a_\mathrm{bg}}{m}
  \nonumber\\
  &\ \ +(2\pi\hbar)^3 4\pi m
   \int_0^\infty dp\,
   |\langle0|V_\mathrm{bg}|\phi_\mathbf{p}^\mathrm{bg}\rangle|^2
   \exp\left(-i\frac{p^2t}{m\hbar}\right)
   \Bigg].
\end{align}
The contributions from the discrete part of the spectrum may be neglected since all binding energies $E_\lambda$ of $^{23}$Na pairs in the $(F=1,m_F=-1)$ ground state are much larger than any other energy scale introduced through the variation of the laser intensity.
We therefore introduced a slowly varying envelope approximation by consistently neglecting all terms oscillating like $\exp\{i|E_{-1}|t/\hbar\}$ or faster.  

The matrix element of $V_\mathrm{bg}$ is rewritten as a matrix element of the $T$-matrix between plane wave momentum states, $\langle0|V_\mathrm{bg}|\phi_\mathbf{p}^\mathrm{bg}\rangle=
\langle0|T_\mathrm{bg}(p^2/m)|\mathbf{p}\rangle$.
For the low scattering energies in a condensate only the partial wave with vanishing angular momentum contributes such that this matrix element only depends on the modulus of $\mathbf{p}$ such that the angular integral could be evaluated when deriving Eq.~(\ref{eq:hbgspec}).
Finally, $h_\mathrm{bg}(t)$ was determined by applying the separable parametrization introduced in Sect.~\ref{app:seppotHom}, i.e.~Eq.~(\ref{eq:TMatbgSepPot}) for the $T$-matrix, and thus expressing the matrix element in Eq.~(\ref{eq:hbgspec}) in terms of the expectation values in Eq.~(\ref{eq:G0Eab}).

In an analogous way, we determined the functions $u_\mathrm{bg}$ and $\tilde{h}_\mathrm{bg}$.
Given these functions, $u_{11}$ and $u_{21}$ were calculated iteratively according to the Volterra-type integral equation (\ref{eq:IntEqforu11and21}), for a particular time profile $I(t)$ of the laser intensity.
Both functions are defined over the triangular plane $t_0\le\tau\le t\le t_\mathrm{max}$, where $t_0$ and $t_\mathrm{max}$ mark the beginning and the end of the laser pulse.
At the end, $h(t,\tau)$ is determined by use of Eq.~(\ref{eq:h11}).

\section{Eigenstates of a non-Hermitian operator}
\label{app:EigenstatesNHOP}
\noindent
The two channel Hamiltonian, Eqs.~(\ref{eq:H2B0nu}), (\ref{eq:V2Bnu}), describing an unstable molecular state $|\phi_\nu\rangle$ in the closed channel coupled to the atoms scattering in the background channel potential $V_\mathrm{bg}$ is non-Hermitian due to the finite lifetime of $|\phi_\nu\rangle$.
Thus the corresponding time evolution operator is non-unitary, i.e.~the norm of any physical eigenstate of the Hamiltonian may decrease in time.

In this appendix we shall, in Subsect.~\ref{app:RuLEigenstates}, recall the basic properties of the eigenstates of non-Hermitian operators.
In Subsect.~\ref{sec:adiablevelsTrap} we shall discuss in further detail the dressed states, i.e.~the spectrum of eigenstates of the Hamiltonian in Eqs.~(\ref{eq:H2B0nu}), (\ref{eq:V2Bnu}) close to threshold.

\label{app:RuLEigenstates}
A suitable approximation method to describe the temporal behaviour of a number of discrete, partially degenerate states, which decay, induced by a weak perturbation, into a range of states with a continuous energy spectrum, relates back to the work of Weisskopf and Wigner \cite{WignerWeisskopf30}.
For not too small and not too large evolution times this approximation leads to the usual exponential decay of the energy eigenstates, described by a decay width contributing as an imaginary part to the energies:
%
\begin{align}
  \label{eq:ComplexEnergies}
  E_j
  &=\epsilon_j-\frac{i}{2}\gamma_j,\quad j=1,2,...
\end{align}
The Hamiltonian may then be expressed in terms of its eigenstates as
%
\begin{align}
  \label{eq:NHHamUnpertBas}
  H
  &= \sum_j |\phi_j\rangle E_j \langle\tilde{\phi}_j|,
\end{align}
where $|\phi_i\rangle$ and $\langle\tilde{\phi}_i|$ are the right and left eigenstates of $H$ to the \emph{same} eigenvalue, respectively, which are defined by the relations
\begin{align}
  \label{eq:reeq}
  H|\phi_j\rangle 
  &=E_j|\phi_j\rangle,
  \\
  \label{eq:leeq}
  \langle\tilde{\phi}_j|H
  &=\langle\tilde{\phi}_j|E_j.
\end{align}
Note that $|\phi_i\rangle$ is, in general, \emph{not} transformed into $\langle\tilde{\phi}_i|$ by a Hermitian conjugation.
The left and right eigenstates form sets of eigenvectors which are dual to each other in the sense that they may be orthonormalized by
\begin{align}
  \label{eq:dualnorm}
  \langle\tilde{\phi}_i|\phi_j\rangle
  &= \delta_{ij}.
\end{align}
Furthermore, the absolute norm of each eigenstate may be fixed at a certain time, by requiring
\begin{align}
  \label{eq:rightnorm}
  \langle\phi_j|\phi_j\rangle
  &= 1.
\end{align}
While the occupation probability $\langle\phi_j(t)|\phi_j(t)\rangle$ exponentially decays in time in the Schr\"odinger picture, the dual norm $\langle\tilde{\phi}_j(t)|\phi_j(t)\rangle$ remains equal to one.
Hence, the normalizations (\ref{eq:dualnorm}) and (\ref{eq:rightnorm}) give the states a physical meaning \emph{and} provide them with properties which are necessary for basis set expansions.
For such expansions one introduces the quasi projectors
\begin{align}
  \label{eq:QuasiP}
  P_j
  &= |\phi_j\rangle \langle\tilde{\phi}_j|,
\end{align}
with the required properties $P_iP_j=\delta_{ij}P_j$ and $\mathrm{Tr} P_j=1$.

In the case that the eigenstates of $H$ with complex energies are coupled to each other by a Hermitian operator, the resulting new right energy eigenstates will in general \emph{not} be orthogonal to each other even if the uncoupled states happened to be so.
However, a dual basis set of left eigenstates may be defined which are orthonormal to the set of right eigenstates.

In the situation considered in this article, a single unstable molecular state is coupled to the spectrum of eigenstates of the background Hamiltonian consisting of a discrete subset of bound states and a continuous part of scattering states.\\

\section{Multichannel formalism of the microscopic many body theory}
\label{sec:multichannel_mb}
Consider a many body ensemble of atoms which may be in a defined number of internal states and collide through the corresponding binary scattering channels.
We label the different internal states by Greek indices.
In the case of photoassociation in sodium the possible asymptotic internal states will be characterized mainly by the electron-nucleus relative angular momenta as well as the total atomic angular momenta of the collision partners.

The photoassociation laser is treated classically, and any coupling of an atom to the continuum of photon modes leading to spontaneous emission is taken into account, in the Wigner-Weisskopf approximation, as a finite width of the relevant atomic energy levels.
The many body Hamiltonian of the atoms may be written as
\begin{widetext}
\begin{align}
  \label{eq:HMBMulCh}
  H(t)
  &= \sum_\kappa\int\,d\mathbf{x}\,
     \psi_\kappa^\dagger(\mathbf{x})
     H_\kappa^\mathrm{1B}(\mathbf{x})
     \psi_\kappa(\mathbf{x})
    +\frac{1}{2}\sum_{\kappa,\lambda,\mu,\nu}
     \int\,d\mathbf{x}d\mathbf{y}\,
     \psi_\kappa^\dagger(\mathbf{x})\psi_\lambda^\dagger(\mathbf{y})
     V_{\kappa\lambda,\mu\nu}(\mathbf{x}-\mathbf{y},t)
     \psi_\mu(\mathbf{x})\psi_\nu(\mathbf{x}).
\end{align}
\end{widetext}
The operators for creation and annihilation of a single atom satisfy the bosonic commutation relations
\begin{align}
  \label{eq:commrels}
  [\psi_\kappa(\mathbf{x}),\psi_\lambda^\dagger(\mathbf{x})]
  &= \delta_{\kappa\lambda}\delta^{(3)}(\mathbf{x}-\mathbf{y}),
  \\
  [\psi_\kappa(\mathbf{x}),\psi_\lambda(\mathbf{x})]
  &= 0.  
\end{align}
The one body Hamiltonian accounts for the kinetic energy of an atom, its energy inside an external trap, and the relative internal energy relative to some arbitrary zero: 
\begin{align}
  \label{eq:H1Bkappa}
  H_\kappa^\mathrm{1B}(\mathbf{x})
  &= -\frac{\hbar^2\nabla^2}{2m} 
     + V_\mathrm{trap}(\mathbf{x}) 
     + E_\kappa^\mathrm{int}.
\end{align}
The coupling functions $V_{\kappa\lambda,\mu\nu}(\mathbf{x},t)$ in the interaction term in Eq.~(\ref{eq:HMBMulCh}) contain all information about the incoming and outgoing microscopic scattering channels which may be labelled by the pairs of asymptotic internal quantum numbers $\{\mu\nu\}$ and $\{\kappa\lambda\}$, respectively.
For the case of photoassociation these are the open background and resonantly coupled closed channels as well as the laser induced coupling which depend on the laser parameters and the Franck-Condon overlaps.
The Hamiltonian includes an explicit time dependence which enters through the temporally variable laser couplings.

The theory is formulated in the Schr\"odinger picture, where the density matrix depends on time.
All physical quantities describing the many body properties are expressed through expectation values of normal ordered products of the time independent field operators with respect to the state at time $t$, denoted as $\langle\cdot\rangle_t$.
For the purpose of this paper we will consider, besides the mean field 
\begin{align}
  \label{eq:Psikappa}
  \Psi_\kappa(\mathbf{x},t)
  &= \langle\psi_\kappa(\mathbf{x})\rangle_t,
\end{align}
the thermal and anomalous density correlation functions $\langle\psi_\lambda^\dagger(\mathbf{y})\psi_\kappa(\mathbf{x})\rangle_t$ and $\langle\psi_\lambda(\mathbf{y})\psi_\kappa(\mathbf{x})\rangle_t$, respectively.

The first characteristic of the theory is a rewriting of the infinite hierarchy of non-linear coupled dynamical equations resulting for these correlation functions from the Hamiltonian (\ref{eq:HMBMulCh}) in terms of noncommutative cumulants \cite{KB02,KGB03}.
The cumulants to be considered in the following are the mean field $\Psi_\kappa(\mathbf{x},t)$, Eq.~(\ref{eq:Psikappa}), the pair function
\begin{align}
  \label{eq:Phikl}
  \Phi_{\kappa\lambda}(\mathbf{x},\mathbf{y},t)
  &= \langle\psi_\lambda(\mathbf{y})\psi_\kappa(\mathbf{x})\rangle_t
     - \Psi_\kappa(\mathbf{x},t)\Psi_\lambda(\mathbf{y},t),
\end{align}
and the density function of the non-condensed fraction:
\begin{align}
  \label{eq:Gammakl}
  \Gamma_{\kappa\lambda}(\mathbf{x},\mathbf{y},t)
  &= \langle\psi_\lambda^\dagger(\mathbf{y})\psi_\kappa(\mathbf{x})\rangle_t
     - \Psi_\kappa(\mathbf{x},t)\Psi_\lambda^*(\mathbf{y},t),
\end{align}

The second characteristic of the approach is a consistent truncation of the system of equations \cite{KB02,KGB03}.
Within the leading order approximation which we will apply in this work the dynamics is determined solely by the equations
\begin{widetext}
\begin{align}
  \label{eq:DynEqPsik}
  i\hbar\frac{\partial}{\partial t}\Psi_\kappa(\mathbf{x},t)
  &= H_\kappa^\mathrm{1B}(\mathbf{x})
     \Psi_\kappa(\mathbf{x},t)
    +\sum_{\lambda,\mu,\nu}
     \int\,d\mathbf{y}\,
     \Psi_\lambda^\dagger(\mathbf{y},t)
     V_{\kappa\lambda,\mu\nu}(\mathbf{x}-\mathbf{y},t)
     \left[\Phi_{\mu\nu}(\mathbf{x},\mathbf{y},t)
           +\Psi_\mu(\mathbf{x},t)\Psi_\nu(\mathbf{x},t)\right],
  \\
  \label{eq:DynEqPhikl}
  i\hbar\frac{\partial}{\partial t}\Phi_{\kappa\lambda}(\mathbf{x},\mathbf{y},t)
  &= \sum_{\mu,\nu}\left[
      H_{\kappa\lambda,\mu\nu}^\mathrm{2B}(\mathbf{x},\mathbf{y},t)
      \Phi_{\mu\nu}(\mathbf{x},\mathbf{y},t)
     +V_{\kappa\lambda,\mu\nu}(\mathbf{x}-\mathbf{y},t)
      \Psi_\mu(\mathbf{x},t)\Psi_\nu(\mathbf{x},t)\right].  
\end{align}
\end{widetext}
where the elements of the Hamiltonian matrix of two interacting atoms are defined as
\begin{align}
  \label{eq:H2Bklmn}
  H_{\kappa\lambda,\mu\nu}^\mathrm{2B}(\mathbf{x},\mathbf{y},t)
  &=\left[H_\kappa^\mathrm{1B}(\mathbf{x})
         +H_\lambda^\mathrm{1B}(\mathbf{y})\right]
    +V_{\kappa\lambda,\mu\nu}(\mathbf{x}-\mathbf{y},t).
\end{align}
The non-condensed atoms are, to this order, described in terms of the pair function as:
\begin{align}
  \label{eq:GammaklitoPhi}
  \Gamma_{\kappa\lambda}(\mathbf{x},\mathbf{y},t)
  &= \sum_\mu\,\int\,d\mathbf{x'}\,
     \Phi_{\kappa\mu}(\mathbf{x},\mathbf{x'},t)
     \Phi_{\lambda\mu}^*(\mathbf{y},\mathbf{x'},t).  
\end{align}
The total number of atoms at time $t$,
\begin{align}
  \label{eq:NtotMulCh}
  N(t)
  &= \sum_\kappa\,\int\,d\mathbf{x}\,\left[
     |\Psi_\kappa(\mathbf{x},t)|^2+
     \Gamma_{\kappa\kappa}(\mathbf{x},\mathbf{x},t)\right]
\end{align}
is conserved in time if the overall system is closed, i.e.~if there is no loss due to, e.g., spontaneous photon emission, to an atomic level not counted to the system.
When considering the system of one open background channel coupled to a closed channel vibrational state, as described in detail in Sect.~\ref{sec:multichannel_2b}, we will find that $N$ is, in general, not conserved due to the nonzero width $\gamma$ of the closed channel state.
Disregarding this loss for a moment, Eqs.~(\ref{eq:GammaklitoPhi}) and (\ref{eq:NtotMulCh}) reveal that the appearance of a non zero pair function, which reflects the buildup of pair correlations, means that atoms have been lost from the condensate to the non-condensed fraction.
These correlated atoms become part of a relatively hot, unequilibrated cloud around the condensate.
In the PA process, where $\gamma>0$, they account for the difference between atoms lost from the condensate and those transferred into ground state molecules through the spontaneous decay process. 
We finally note that inelastic loss phenomena due to e.g.~three body recombination \cite{KB02,Koehler02} are neglected in the leading order approach presented here.

\section{Time evolution of the total atom number}
\label{app:optth}
\noindent
In the absence of decay the total number of atoms $N_\mathrm{tot}(t)=\int d\mathbf{x}\,n_\mathrm{tot}(\mathbf{x},t)$ in the condensate and the excited states, with $n_\mathrm{tot}$ given by Eq.~(\ref{eq:ntot}), is constant in time.
This general property of the microscopic dynamical theory was shown for the one-channel case in Ref.~\cite{KGB03} and generalized for two channels in Ref.~\cite{GKGJT04}.
For photoassociation which involves resonant coupling to an unstable closed channel state, however, the system is no longer closed and the total number of atoms may decrease in time.
In this appendix, we sketch the line of arguments leading to the evolution equation (\ref{eq:dNndt}) for the total atom number.

According to Refs.~\cite{KGB03,GKGJT04} the time derivative of the total number of non-condensed atoms $\Gamma(\mathbf{x},\mathbf{y},t)=\Gamma_\mathrm{bg}(\mathbf{x},\mathbf{x},t)+\Gamma_\mathrm{cl}(\mathbf{x},\mathbf{y},t)$ may be written as
%
\begin{align}
  \label{eq:dGammatotndt}
  \frac{\partial}{\partial t}N_\mathrm{nc}(t)
  &=
  \frac{\partial}{\partial t}\int d\mathbf{x}\,
  \Gamma(\mathbf{x},\mathbf{x},t)
  \nonumber\\
  &= 2i\mathrm{Im}\left[\int d\mathbf{x}d\mathbf{y}
    \Phi^*(\mathbf{x},\mathbf{y},t)\frac{\partial}{\partial t}
    \Phi(\mathbf{x},\mathbf{y},t)\right],
\end{align}
%
where $\Phi=|\mathrm{bg}\rangle\Phi_\mathrm{bg}+|\mathrm{cl}\rangle\Phi_\mathrm{cl}$ includes the background and closed channel pair functions.
Using Eqs.~(\ref{eq:meanfield}) and (\ref{eq:pairfunction}) for the time evolution of the mean field and pair function one finds: 
\begin{align}
  \label{eq:dNtotndtitoPhi}
  \frac{\partial}{\partial t}(N_\mathrm{c}(t)+N_\mathrm{nc}(t))
  &= \langle\Phi(t)|[H_\mathrm{2B}(t)-H_\mathrm{2B}^\dagger(t)]|\Phi(t)\rangle,
\end{align}
where $N_\mathrm{c}=\int d\mathbf{x}\,|\Psi(\mathrm{x})|^2$ is the number of condensed atoms and $|\Phi(t)\rangle$ denotes in a basis independent way the pair function $\Phi$ (cf.~Sect.~\ref{sec:inicor}).
For obtaining Eq.~(\ref{eq:dNtotndtitoPhi}) it is essential that there is no direct decay from the background channel, i.e.~$V_\mathrm{bg}^\dagger=V_\mathrm{bg}$, and that a condensate only exists for atoms in the internal ground state, as discussed in Sect.~\ref{sec:2chdyneq}.

The term on the right of Eq.~(\ref{eq:dNtotndtitoPhi}) is non-zero due to the imaginary part of the closed channel diagonal element of the Hamiltonian in Eq.~(\ref{eq:H2B0I}).
Hence, inserting Eq.~(\ref{eq:H2B0I}), we obtain 
\begin{align}
  \label{eq:dNtotndtitoPhiandphinu}
  \frac{\partial}{\partial t}(N_\mathrm{c}(t)+N_\mathrm{nc}(t))
  &= -\frac{\gamma}{\hbar}|\langle\mathrm{cl},\phi_\nu|\Phi(t)\rangle|^2.
\end{align}
Here, we have once more used that the hermitian conjugate of the \emph{time independent} right eigenstate $|\phi_\nu\rangle$ is equivalent to the left eigenstate $\langle\tilde\phi_\nu|$.
Finally, solving Eq.~(\ref{eq:pairfunction}) one obtains an expression for $\Phi_\mathrm{cl}(\mathrm{x},\mathrm{y},t)$ which by insertion into Eq.~(\ref{eq:dNtotndtitoPhiandphinu}) yields the desired relation (\ref{eq:dNndt}).
Thereby, the zero momentum plane wave state results from the approximation \cite{KGB03,GKGJT04} that the mean field $\Psi(\mathbf{x},t)$ is constant on the scale of the variation of the background potential $V_\mathrm{bg}(r)$ and the closed channel state $\phi_\nu(r)$.

\end{appendix}

%


\begin{thebibliography}{99}
\bibitem{fn:poleapprox}
Alternatively one may keep the full closed channel Hamiltonian $H_\mathrm{cl}$ until this step of the derivation, and now choose the pole approximation $G_\mathrm{cl}(z,\Delta)$ of the Green's function $G_\mathrm{cl}(z)$ as given in Eq.~(\ref{eq:Gclofz}).
\bibitem{fn:Rb85case}
Close to a Feshbach resonance where the scattering length $a$ is large and positive, $\bar{a}$ represents a small correction to $a$, and the binding energy is well approximated by $E_\mathrm{b}=-\hbar^2/m(a-\bar{a})^2$ as may be seen, e.g., in Fig.~14 of Ref.~\cite{KGB03}.
\bibitem{fn:reducedmassinholength}
Note that in Ref.~\cite{Bolda02} the oscillator length was defined in terms of the reduced mass $\mu=m/2$.
\end{thebibliography}
\end{document}